\documentclass{article}

\usepackage{svg}
\usepackage{setspace}
\usepackage{xcolor}
\usepackage{hyperref}
\usepackage{caption}
\usepackage{listings}
\usepackage{morefloats}
\usepackage{float}
\usepackage{fancyhdr}
\usepackage{mathtools}
\usepackage{natbib}
\usepackage{bm}
\usepackage{bbm}
\usepackage{amsfonts}
\usepackage{tabularx}
\usepackage{booktabs}
\usepackage[margin=1.05in]{geometry}
\usepackage[toc,page]{appendix}

\usepackage{makecell}

\definecolor{upennblue}{RGB}{1,31,91}
\definecolor{upennred}{RGB}{153,0,0}

\newcommand{\E}[1]{\text{E}\left[#1\right]}
\newcommand{\Eind}[2]{\text{E}_{#1}\left[#2\right]}
\newcommand{\biground}[1]{\left(#1\right)}
\newcommand{\x}{\mathbf{x}}
\newcommand{\nn}{\mathcal{N}_\rho}

\usepackage{xargs}

\hypersetup{colorlinks=true, urlcolor=upennblue, citecolor=upennblue, 
linkcolor = upennred,
pdftitle = {AzinovicZemlicka_2023_EconomicsInspiredNeuralNetworksWithStabilizingHomotopies}}

\title{Economics-Inspired Neural Networks with Stabilizing Homotopies\thanks{Preliminary and incomplete, comments are welcome. We are grateful to Jes\'us Fern\'andez-Villaverde, Dirk Krueger, Marek Kapička, Simon Scheidegger, and Yucheng Yang for helpful comments. We are also grateful to the Economics Department at the University of Pennsylvania for their kind hospitality.}}

\author{Marlon Azinovic\footnote{Email: azinovic@sas.upenn.edu. Azinovic acknowledges support from the Swiss National Science Foundation under grant P500PS\_210799 ``\emph{Macro-Finance with Heterogeneous Households}''.}\\ University of Pennsylvania \and Jan Žemlička\footnote{Email: jan.zemlicka@cerge-ei.cz. This paper was supported by the grant from the Charles University Grant Agency under project No. 192822 ``\emph{Advanced Neural Networks Architectures for Solving Heterogeneous Agent Models}''.} \\ CERGE-EI\footnote{CERGE-EI, a joint workplace of Charles University and the Economics Institute of the Czech Academy of Sciences, Politickych veznu 7, 111 21 Prague, Czech Republic}}

\date{First version: March 12, 2023\\
This version: \today
}

\begin{document}
\maketitle

\begin{abstract}
\singlespacing
Contemporary deep learning based solution methods used to compute approximate equilibria of high-dimensional dynamic stochastic economic models are often faced with two pain points.
The first problem is that the loss function typically encodes a diverse set of equilibrium conditions, such as market clearing and households' or firms' optimality conditions.
Hence the training algorithm trades off errors between those --- potentially very different --- equilibrium conditions.
This renders the interpretation of the remaining errors challenging.
The second problem is that portfolio choice in models with multiple assets is only pinned down for low errors in the corresponding equilibrium conditions.
In the beginning of training, this can lead to fluctuating policies for different assets, which hampers the training process.
To alleviate these issues, we propose two complementary innovations. First, we introduce \emph{Market Clearing Layers}, a neural network architecture that automatically enforces all the market clearing conditions and borrowing constraints in the economy. Encoding economic constraints into the neural network architecture reduces the number of terms in the loss function and enhances the interpretability of the remaining equilibrium errors. 
Furthermore, we present a homotopy algorithm for solving portfolio choice problems with multiple assets, which ameliorates numerical instabilities arising in the context of deep learning. To illustrate our method we solve an overlapping generations model with two permanent risk aversion types, three distinct assets, and aggregate shocks.
\end{abstract}
\medskip


\clearpage

\section{Introduction}
A rapidly growing literature \citep[see, \emph{e.g.}][]{duarte2018machine, valaitis2021machine, azinovic2022deep, kahou2021exploiting, han2021deepham, maliar2021deep, kase2022estimating, duarte2021simple, folini2021climate, bretscher2022ricardian} uses deep neural networks to approximate price, policy and value functions in economic models. To find good approximations, the learning algorithm often aims directly at minimizing the errors in the equilibrium conditions, such as Euler equations, market clearing conditions, and budget constraints, on simulated paths of the economy.\footnote{Examples of deep learning algorithm which directly minimize errors in the equilibrium conditions are \cite{azinovic2022deep, maliar2021deep, kahou2021exploiting, kase2022estimating}.} Since the equilibrium conditions will not be fulfilled exactly, the researcher has to, implicitly, trade-off errors between different equilibrium conditions.
This can be challenging, especially for equilibrium conditions of different types and units.

We propose a new architecture for deep neural networks, which we call market clearing layers, that, by construction, enforces exact market clearing conditions. Encoding known properties, coming from the economic model, directly into the architecture has the advantage that it reduces the search space in the space of neural network approximators, to only the subset consistent with market clearing.
Hence the neural network does not need to learn a property, we know ex-ante. Furthermore, the number of different error terms in the loss function is reduced, and hence a given loss value is easier to interpret.
Next to a simple and easily implementable way to ensure market clearing, we also introduce an implicit layer, which additionally ensures that borrowing constraints are always satisfied.
To do so, the last layer of the network solves a quadratic program to obtain adjustments to the layer inputs, with computationally desirable properties. We are not the first to leverage economic insights for the design of neural network architectures. \cite{kahou2021exploiting} and \cite{han2021deepham} encode a symmetry property into the neural network architecture. \cite{han2021deepham} and \cite{azinovic2022deep} use suitable activation functions in the output layers to ensure borrowing constraints are satisfied.

Additionally, we propose a homotopy method to stabilize the training progress of deep learning based solution methods for economic models with multiple assets.
Homotopy methods are a class of methods, which start by solving a simple problem and then gradually transform the simple problem into the harder problem of interest.
In our case, the simple problem to start from is an economic model with a single asset.
The hard problem of interest is a model with multiple assets.
Multiple assets pose a problem for deep learning based solution methods since a precise solution is required for agents to correctly distinguish between multiple assets.
Especially during the beginning of training, when the policies are not yet precisely approximated, this can lead to erratic policies with spurious bang-bang allocations to different assets, which may lead to a failure of the training procedure.
To address this, we propose to firstly solve a single-asset model, by setting the supply and short-sale constraints of the remaining assets to zero.
Once the single asset model is solved, we keep training the model and slowly increase the supply of the second asset. We proceed this way until the full model is solved.
The idea to use homotopy methods to solve for equilibria in economic models has been used, for example, in \cite{schmedders_1998}, where a homotopy is used to gradually lift bounds on asset sales.
The idea to introduce small amounts of an additional asset is reminiscent of the idea in \cite{juddguu_2000}, who first solve a model with deterministic asset returns and then uses perturbation methods to introduce small amounts of uncertainty into the asset returns.

The two contributions of this paper go hand in hand. Since the homotopy method we are proposing, slowly and one after the other, increases the supply of additional assets in the economy, it naturally leverages the market clearing neural network architectures.

To the best of our knowledge, we are the first to study market clearing architectures for neural networks and to lay out a homotopy algorithm to step-wise introduce multiple assets into a deep learning based solution method.

\section{Market Clearing Layers\label{sec:mcl}}
To introduce market clearing layers, we first consider a neural network approximating policy and price functions for a single asset and a finite number of agents, abstracting from the rest of the economy, such as the households' objective functions.
In section \ref{sec:simple_example} we show how market clearing layers can be applied in a concrete economic model.

Consider an economic model, with a set of finitely many agents, indexed with $i\in \mathcal{I}:=\{1, \dots, H\}$ and associated population weights $\mu_i$.
In an overlapping generations model, for example, $i$ could index a specific age group and $\mu_i$ could denote its size.\footnote{In an economy with a continuum of agents, $i$ could index a specific combination of shocks and asset holdings, and $\mu_i$ would denote the mass of agents at a specific histogram point.}
Let $\x \in \mathbb{R}^N$ denote the state of the economy.
Assume that agents can save in a single asset $b$ with price $p^b$ and aggregate supply $B$. Let $b^i(\x)$ denote the policy function of agent $i$.
Market clearing requires that the total asset demand by households adds up to the total asset supply:
\begin{align}
\sum_{i=1}^H \mu_i b^i(\x) = B.
\end{align}
We are interested in using a neural network to approximate the households' asset policies.
Let 
\begin{align}
\nn : \mathbb{R}^N \rightarrow \mathbb{R}^H\times \mathbb{R}_{> 0}, \nn(\x) = \begin{pmatrix}
\tilde{b}^1(\x)\\
\tilde{b}^2(\x)\\
\dots\\
\tilde{b}^H(\x)\\
\hat{p}^b(\x)
\end{pmatrix}
\end{align}
denote a neural network with parameters $\rho$ that maps the state of the economy to an approximation of the household savings choices and the price of the asset.
The goal of this paper is to devise a neural network architecture that, by design, ensures that market clearing is satisfied while also retaining desirable computational properties.

\subsection{Simple Adjustment for Pure Market Clearing\label{sec:simple}}
Let the aggregate asset demand, which is implied by the households' policies $\tilde{b}^i(\x)$, be denoted with
\begin{align}
B^H(\x) := \left(\sum_{i=1}^{H} \mu_i \tilde{b}^i(\x)\right).
\end{align}
The corresponding excess demand is given by
\begin{align}
\Delta B(\x) = B^H(\x) - B.
\end{align}
We would like to adjust the policies $\tilde{b}^i(\x)$, such that they are consistent with market clearing, \emph{i.e.} such that the excess demand is zero.
Let $b^i(\x)$ denote the adjusted policy functions.
There are multiple ways to adjust the policies $\tilde{b}^i$ to make up for the excess demand $\Delta B(\x)$.
For example, one way would be to rescale the policies to obtain new policies
\begin{align}
\forall i\in \mathcal{I}:~ b^i(\x) = \tilde{b}^i(\x)\frac{B}{B^H(\x)}.
\end{align}
However, in general this algorithm would not be convenient, for example in settings when $B^H(\x) = 0$.
Another way would be to adjust the policy of only a single household, for example
\begin{align}
j\in \mathcal{I}: b^j(\x) &= \tilde{b}^j(\x) - \frac{\Delta B(\x)}{\mu^j}\\
\forall i\in \mathcal{I}, i\neq j: b^i(\x) &= \tilde{b}^i(\x).
\end{align}
This way would load all the excess demand onto a single agent and cause an asymmetry in the accuracy of the policies across different households.

We propose the solution to the following problem as a desirable adjustment mechanism:
\begin{align}
\{b^i(\x)\}_{i\in \mathcal{I}} &=\arg\min \sum_{i\in \mathcal{I}}\frac{1}{2}\mu_i (b^i(\x) - \tilde{b}^i(\x))^2 \nonumber\\
\text{subject to}: \sum_{i\in \mathcal{I}}\mu_i b^i(\x) &= B.\label{eq:simpleadj}
\end{align}
The problem has the simple solution that all households policy is adjusted by an equal amount
\begin{align}
\forall i \in \mathcal{I}: b^i(\x) = \tilde{b}^i(\x) - \Delta B(\x).\label{eq:formula_simpleadj}
\end{align}
Next to avoiding a distortion in the symmetry between the Euler equations across households this formulation has the advantage that it also works for assets in zero net supply as well as for assets where the net supply is state-dependent.\footnote{The latter is true for the previously mentioned to formulations as well.}

Despite its computational simplicity, a disadvantage of this approach is that, while the adjusted policies are consistent with market clearing, they are not necessarily consistent with borrowing constraints.
A softplus activation function, for example, could ensure that $\tilde{b}^i(\x)$ are non-negative, however this would not imply that the adjusted, market clearing, $b^i(\x)$ are non-negative as well.
Ideally, we would like the adjustment to be able to ensure that both, market clearing and borrowing constraints, are always satisfied.
We address this point in the next section.

\subsection{Implicit Layer to Encode Market Clearing and Borrowing Constraints \label{sec:marketclearing_const}}
The market clearing adjustment described in the previous section does not ensure that borrowing constraints are satisfied.
One way to deal with this would be to separately predict Karush-Kuhn-Tucker (KKT) multipliers and to include the KKT conditions into the loss function. An alternative would be to use the Fischer-Burmeister equation \citep[see][]{jiang_1999}, as for in example \cite{maliar2021deep}, to encode the Euler equation error and the violation of the borrowing constraint into a single error term.
We propose a modification to the market clearing mechanism described in the previously in section \ref{sec:simple}, which simultaneously ensures market clearing and that the borrowing constraints are satisfied, by adding the borrowing constraint to the above minimization problem:
\begin{align}
\{b^i(\x)\}_{i\in \mathcal{I}} &=\arg\min \sum_{i\in \mathcal{I}}\frac{1}{2}\mu_i (b^i(\x) - \tilde{b}^i(\x))^2 \nonumber \\
\text{subject to}: \sum_{i\in \mathcal{I}}\mu_i b^i(\x) &= B \nonumber \\
b^i(\x) &\geq \underline{b}^i(\x) \label{eq:implicitprob}
\end{align}
where $\underline{b}^i(\x)$ denotes the borrowing limit of agent $i$.
The solution to this problem can be obtained with solvers for box-constrained Quadratic Programs, which are meanwhile implemented in modern deep learning libraries, such as JAX, and provide efficiently computed derivatives for the use in backpropagation algorithms.\footnote{We used the \texttt{BoxOSQP} solver from the \texttt{jaxopt} library: \url{https://jaxopt.github.io/stable/quadratic_programming.html}.}
The draw-back of these algorithms is that they substantially slow down the training process and are often programmed for a general class of Quadratic Programs and do not exploit symmetries, which are often present in economic models.

Consider, for example, a borrowing constraint at zero for all agents: $\forall{i}:~b^i(\x)\geq 0$.
In this case we can ensure that $\tilde{b}^i\geq 0$ by using a softplus activation function in the output layer of the neural network.
If the excess demand is negative, \emph{i.e.} $\Delta B(\x)<0$, the solution to the constrained quadratic program coincides with the simple solution described in section \ref{sec:simple}.
The adjusted asset demand of all agents is given by
$b^i(\x) = \tilde{b}^i(\x) - \Delta B(\x)$ and no constrained is violated.
When the excess demand is positive, all asset choices $\tilde{b}^j$ below a threshold value $\tilde{b}^{\text{threshold}}$ are adjusted to lie on the borrowing constraint and all remaining asset holdings are adjusted by the same amount such that markets clear.
Let $\mathcal{J}$ denote the set of households $j$ with $\tilde{b}^j < \tilde{b}^{\text{threshold}}$.
The solution to the constrained quadratic program is given by
\begin{align}
b^i = 
\begin{cases}
0 &\text{for }i \in \mathcal{J}\\
\tilde{b}^i - \underbrace{(\Delta B(\x) - \sum_{j\in \mathcal{J}}\mu_j(\tilde{b}^j - 0))}_{\text{remaining excess demand}}&\text{for }i \in \mathcal{I} \setminus\mathcal{J}
\end{cases}
\end{align}
Solving the constrained quadratic program hence simplifies to finding a single number $\tilde{b}^{\text{threshold}}$.
Since $\tilde{b}^{\text{threshold}} \in [\min_{i}\tilde{b}^i, \max_i\tilde{b}^i+\epsilon] $, for $\epsilon > 0$,  it can be efficiently computed with a few bisection steps.
For our examples, this turned out to be faster than using the general solver for constrained quadratic programs.
\subsubsection{Policy updating for falsely constrained agents\label{sec:vanishingpass}}
One problem, when ensuring the borrowing constraint as described above, is that for constrained households, \emph{i.e.} households with $\tilde{b}^i < \tilde{b}^{\text{threshold}}$ and $b^i = 0$, an infinitesimal change in the neural network output $\tilde{b}^i$ does not influence the adjusted policy $b^i$.
For a loss function purely based on the post-adjustment prediction $b^i$, this is a problem if an agent is predicted to be constrained, though it should not be. 
The derivative of the loss function with respect to the pre-adjustment prediction $\tilde{b}^i$ would be zero and the gradient would hence not help to improve the corresponding parameters in the neural network.\footnote{A similar problem could arise in models where the last layer has a softplus activation when the pre-activated value is very negative. While the softplus activation would still guarantee a positive derivative, the derivative could be vanishingly small.}
To address this issue, we provide a way to propagate the signal from the loss function to the neural network, so that the gradient descent step will lead to improved neural network parameters, which increase the prediction for $\tilde{b}^i$ in such cases.
For clarity, we detail it in context of the simple example below in section \ref{sec:simple_example}.
While our solution allows us to retain the signal while enforcing the borrowing constraints by design, we admit that the solution is based on additional terms in the loss function, which is unfortunate given the very goal of this paper.
Another solution, of course, is to follow the simple adjustment described above in section \ref{sec:simple}. Using reformulations like the Fischer-Burmeister equation \citep[see][]{jiang_1999},\footnote{The Fischer-Burmeister function is $\psi(x, y) := x + y - \sqrt{x^2+y^2}$. While the Fischer-Burmeister function is differentiable, its roots satisfy $x\geq 0$, $y \geq 0$ and $xy = 0$.} or the Garcia Zangwill-trick \citep[see][]{garciazangwill_1981, juddetal_2000}, the KKT conditions can be summarized by a single equation and no additional terms in the loss function are required.

\subsection{Choosing Between the Two Market-Clearing Algorithms}
In sections \ref{sec:simple} and \ref{sec:marketclearing_const} we laid out two ways to ensure that market clearing is satisfied exactly.
The simple algorithm described in section \ref{sec:simple} has the advantage that we can obtain the solution to problem \eqref{eq:simpleadj} in closed form, as given in equation \eqref{eq:formula_simpleadj}.
The market clearing policies can hence be computed with minimal computational complexity.
The disadvantage is that the adjusted policies may violate the borrowing constraints. Consequently, the borrowing constraints must be encoded in the loss function, for example by using the Fischer-Burmeister equation \citep[see][]{jiang_1999}.

The solver-based adjustment described in section \ref{sec:marketclearing_const} has the advantage that the solution to the problem \eqref{eq:implicitprob} ensures that, additionally to market clearing, all borrowing constraints are always satisfied.
The disadvantage is that a closed form solution for constrained quadratic programs is not available and we therefore need to invoke a solver, which slows down the training.
Furthermore, the solver-based adjustment faces the problem that we need to add auxiliary terms to the loss-function in order to ensure that the neural network weights are updated if an agent is falsely predicted to be constrained, as described in section \ref{sec:vanishingpass}.

Which of the two market clearing algorithm is better may depend on the model at hand and on whether a strict enforcement of the borrowing constraints adds stability to the training algorithm.
When the borrowing constraints can be conveniently encoded using the Fisher-Burmeister equation, we found that the advantage of ensuring the borrowing constraints by design is not large enough to make up for the loss in speed resulting from the invocation of a quadratic solver each time the neural network is evaluated. 

\subsection{Learning algorithm\label{sec:learn_alg}} 
We use a version of the Deep Equilibrium Nets method developed by \cite{azinovic2022deep}. Following the Deep Equilibrium Nets algorithm, we proceed in three steps. First, we parameterize the equilibrium functions of the economy (i.e. value, policies, prices, etc) using a deep neural network. Second, we write down a loss function, which is defined as a mean squared residual of model equations, which characterize the equilibrium functions, over a batch of states. Third, we update network weights by minimizing the loss function along simulated paths of the economy.

Because the loss function includes all the equilibrium conditions of the economy, achieving a sufficiently low level of loss over some region of the state space indicates that the neural network learned how to construct an approximate model solution over that region. By minimizing loss function over states obtained from a long simulation of the economy we make sure that the neural network learns to approximate the model solution on the approximate ergodic set of the economy. As discussed in \cite{judd2011numerically}, drawing states from the approximate ergodic set instead of sampling from some of its enclosing hypercubes allows for a drastic reduction of the computational complexity of obtaining an approximate solution.\footnote{Sparse grids \citep[see][]{kruegerkubler_2004} and adaptive sparse grids \citep[see][]{brumm2017using}, for example, require a hyper-cubic domain.}

Relative to the original Deep Equilibrium Nets algorithm, our algorithm differs along two dimensions. First, our neural networks feature Market Clearing Layers, hence their outputs automatically satisfy market clearing conditions. Therefore, market clearing conditions do not have to be included in the loss function. Moreover, the states we obtain from the simulation are always consistent with market clearing.
Second, we use a modified simulation algorithm. Instead of simulating several paths of the economy for many periods before retraining the neural network, we simulate a large number of states for only one period forward before updating the network weights by training on those newly simulated states.\footnote{Results shown by \cite{zemlicka2023cloud} suggest, that this approach improves the stability of the training process.} 

Specifically, we simulate a number of parallel trajectories for one period forward.
Let $N^{\text{trajectories}}$ denote the number of parallel trajectories.
To simulate the model forward, we use the policies as approximated by the neural network and draw random exogenous shocks according to their transition probabilities.
In each simulation step, we hence obtain $N^{\text{trajectories}}$ new states.
These newly simulated state constitute our dataset to update the approximated policies by training the neural network on the dataset to minimize the loss function.
Let $N^{\text{epochs}}$ denote the number of epochs we train on each simulated dataset, and let $N^{\text{mini-batch}}$ denotes the mini-batch size.
An epoch refers to a single training pass through the entire dataset.\footnote{See \cite{goodfellow2016deep} for a general introduction to deep learning.} 
The mini-batch size refers to the number of data-point used for each step of stochastic gradient descent.\footnote{We use the Adam optimizer \citep[][]{kingma2014adam}, which is a variant of stochastic gradiant descent and the de-facto standard in the deep learning literature.}
Hence, each simulated dataset of $N^{\text{trajectories}}$ new states is used for $\frac{N^{\text{trajectories}}N^{\text{epochs}}}{N^{\text{mini-batch}}}$ steps of stochastic gradient descent.
After training the neural network, we simulate a new dataset.
We refer to one simulation of new training data, together with subsequent training of the neural network on that data, as one episode.
Let $N^{\text{episodes}}$ denote the total number of episodes used to train the neural network.
Our simulation procedure starts out with $N^{\text{trajectories}}$ copies of a single feasible state of the economy.
As the policies improve and as we simulate the model forward, the set of states, generated by simulation, converges toward an approximation of the ergodic set of states of the economy.

In order to evaluate the accuracy of our approximate solution, we simulate the model forward without retraining the neural network parameters and evaluate the equilibrium conditions on the newly generated set of states.

\subsection{Simple Example\label{sec:simple_example}}
We illustrate the market clearing mechanism with an overlapping generations economy, with a single risk free asset and borrowing constraints.
\subsubsection{Model \label{sec:simple_model}}
Time is discrete.
Every period, a shock $z_t$ realizes, which determines the aggregate labor income.
We assume that $\log(z_t)$ follows an AR(1) process
\begin{equation}
\log(z_t) = \rho \log(z_{t-1}) + \sigma \epsilon_t, \label{eq:ar1}
\end{equation}
where $\rho$ denotes the persistence and $\sigma$ the volatility, and where the innovation $\epsilon_t$ is drawn form a standard normal distribution.

Every period a new cohort of measure one enters the economy and lives deterministically for $H<\infty$ periods.
Households of age $h\in \{1 \dots H\}$ receive an exogenous income $y^h_t= z_t y^h$ which is given by an age-specific share $y^h$ of aggregate labor income $z_t$.
Households rent housing services with fixed supply $H^r$, consume and can trade a risk free one-period bond with fixed supply $B$ subject to borrowing constraints.
Let $b_t^h$ denote the bond holdings of age group $h$ in the beginning of period $t$ and let $\underline{b}$ denote the exogenous borrowing constraint: $b_{t+1}^{h+1}\geq \underline{b}$.
We assume that households are born and die without assets, $b_t^1 = b_t^{H+1} = 0$.
To adjust their bond holdings, households have to pay an adjustment cost $\zeta^b\frac{1}{2}(b_{t+1}^{h+1} - b_{t}^{h})^2$.
Households of age $h$ maximize their remaining time-separable discounted expected lifetime utility given by
\begin{align}
\sum_{i = 0}^{H - h}\Eind{t}{\beta^{i}\left(u(c_{t+i}^{h+i}) + \psi^h v(h_{t+i}^{r, h+i})\right)},
\end{align}
where $c_t^h$ denotes the consumption of age group $h$ in period t, $h_t^{r,h}$ denotes the consumption of housing services of age group $h$ in period $t$, and $\beta$ denotes the patience.
We assume constant relative risk aversion utility for both types of consumption with coefficient of relative risk aversion $\gamma$, $u(c)=\frac{c^{1-\gamma}}{1-\gamma}$ and $v(h^r)=\frac{(\underline{h}+ {h^r})^{1-\gamma}}{1-\gamma}$.
The weight for housing services, $\psi^h$, is age dependent.
Households budget constrained is given by
\begin{align}
c_t^h = z_t y^h + b_t^h - p^b_t b_{t+1}^{h+1} - p^b_t \zeta^b\frac{1}{2}(b_{t+1}^{h+1} - b_{t}^{h})^2 - p^{r}_t {h^r}_{t+1}^{h+1},
\end{align}
where $p^b_t$ denotes the price of the bond and $p^{r}_t$ denotes the price for renting housing services.
The Bellman equation corresponding to the households' problem is given by
\begin{align}
V^h(z_t, b_t^h) &= \max_{b_{t+1}^{h+1}, h^{r, h}_{t}}\left\{u(c_t^h) + \psi^h v(h^{r, h}_{t}) + \beta \E{V^{h+1}(z_{t+1}, b_{t+1}^{h+1}}\right\}\\
\text{subject to}:& \nonumber \\
c_t^h &= z_t y^h + b_t^h - b_{t+1}^{h+1} p_t^b - h^{r, h}_{t}p_t^{r}- \zeta^b \frac{1}{2} \biground{b_{t+1}^{h+1} - b_{t}^{h}}^2\\
0 &\leq b_{t+1}^{h+1} - \underline{b} 
\end{align}
The Karush-Kuhn-Tucker conditions for the savings choice of and households' savings choices are given by
\begin{align}
(p_t^b + p_t^b \zeta^b (b_{t+1}^{h+1} - b_{t}^{h})) u'(c_t^h)&=\beta \Eind{t}{u'(c_{t+1}^{h+1})(1 + p^b_{t+1} \zeta^b (b_{t+2}^{h+2} - b_{t+1}^{h+1})) } + \lambda_t^h\\
(b_{t+1}^{h+1} - \underline{b}) &\geq 0\\
\lambda_t^h &\geq 0\\
\lambda_t^h(b_{t+1}^{h+1} - \underline{b}) &= 0.
\end{align}
Similarly, optimality conditions for the intratemporal renting choice are given by
\begin{align}
p^{r}_t u'(c_t^h)&=\psi^h v'(h^{r, h}_t).
\end{align}
We provide details on parameter choices in the appendix \ref{sec:params_single}.
\paragraph{Functional Rational Expectations Equilibrium}
Let $\mathbf{b} := [b^1, \dots, b^H] \in \mathbb{R}^{H}$ denote the distribution of asset holdings across age groups.
Let $\x := [z, \mathbf{b}] \in \mathbb{R}^{H+1}$ denote the state of the economy.
Following \cite{spear_1988} and \cite{kruegerkubler_2004} a functional rational expectations equilibrium is given by a set of $H-1 + H$ policy functions, which map the state of the economy the bond purchases and the consumption of rental services of each age group, together with two price functions, which map the state of the economy to the prices of the bond and rental services, such that the policies and prices are consistent with market clearing as well as the households' optimality conditions. 
\paragraph{Approximating the equilibrium functions with deep neural networks}
Following \cite{azinovic2022deep}, we aim to solve the model by approximating the equilibrium functions with a deep neural network.
Let $\nn : \mathbb{R}^{H+1} \rightarrow \mathbb{R}^{2 H + 1}$ denote a neural network we use to approximate the price and policy functions, such that
\begin{align}
\nn(\x_t) = 
\begin{bmatrix}
\tilde{b}_t^1 \\
\dots \\
\tilde{b}_t^{H-1} \\
\tilde{h}_t^{r, 1} \\
\dots \\
\tilde{h}_t^{r, H-1} \\
\hat{p}^b_t\\
\hat{p}^{r}_t
\end{bmatrix}
\end{align}
In difference to \cite{azinovic2022deep} we now apply the market clearing layers described in section \ref{sec:mcl} to obtain approximations for the households' policies $\hat{b}_t^1, \dots, \hat{b}_t^{H-1}$ and $\hat{h}_t^{r,1}, \dots, \hat{h}_t^{r, H-1}$, which are already consistent with market clearing.
We illustrate both ways, the simple way described in section \ref{sec:simple} and the adjustment described in section \ref{sec:marketclearing_const}, which additionally ensures that the borrowing constraints are satisfied.
Due to the Inada property of the utility from housing services, there is no borrowing constraint for renting services and the two adjustments coincide for renting.

\paragraph{Loss function}
Since our algorithm ensures that market clearing holds exactly for all predicted policy functions, the remaining equilibrium conditions are the households' first order conditions.
As in \cite{azinovic2022deep} we construct a loss function as the means squared error in the remaining equilibrium conditions and the train the neural network to minimize those errors on simulated paths of the economy.
Let $\hat{c}_t^h$ denote the consumption of age group $h$, which the neural network, together with the market clearing layer, predicts for state $\x$.
We use the Fischer-Burmeister function, $\psi(x, y) := x + y -\sqrt{x^2 + y^2}$ to encode the KKT conditions for each age group into a single equation.
Following \cite{judd1998numerical} we rewrite the equations, such that deviations from fullfilling them exactly are interpretable as relative consumption errors.
The resulting errors in the equilibrium conditions for age group $h$ are given by
\begin{align}
\text{err}^{c, h}_\rho(\x_t) &= \psi\biground{\frac{(u')^{-1}\biground{\frac{\beta \Eind{t}{u'(\hat{c}_{t+1}^{h+1})(1 + \hat{p}^b_{t+1}\zeta^b (\hat{b}_{t+2}^{h+2} - \hat{b}_{t+1}^{h+1}))}}{\hat{p}^b_t + \hat{p}^b_t \zeta^b (\hat{b}_{t+1}^{h+1} - \hat{b}_{t}^{h})}}}{\hat{c}_t^h} - 1, \frac{\hat{b}^{h+1}_{t+1} - \underline{b}}{\hat{c}_t^h}} \label{eq:hh_fb_bond}\\
\text{err}^{r, h}_\rho(\x_t) &= \frac{(u')^{-1}\biground{\frac{\psi^h v'(\hat{h}_t^{r, t})}{\hat{p}_t^r}}}{\hat{c}_t^h} - 1.  \label{eq:hh_fb_rent}
\end{align}
Equation \eqref{eq:hh_fb_bond} encodes the errors in the optimality conditions for the choice of saving in the bond, and equation \eqref{eq:hh_fb_rent} encodes the errors in the optimality conditions for the choice of renting housing services.
Let $\mathcal{D}$ denote a set $|\mathcal{D}|$ states.
Our loss function, which encodes all remaining equilibrium conditions, is given by
\begin{align}
l_{\mathbf{\rho}}(\mathcal{D}) := \frac{1}{|\mathcal{D}|}\sum_{\x \in \mathcal{D}}\biground{\frac{1}{H - 1} \sum_{h = 1}^{H - 1} \biground{\text{err}^{c, h}_\rho(\x)}^2 + \frac{1}{H} \sum_{h = 1}^{H} \biground{\text{err}^{r, h}_\rho(\x)}^2}.
\end{align}

\paragraph{Updating the policies of falsely constrained households}
When using our solver based method, there would be no feedback from the loss function to a neural network prediction $\tilde{b}^h$ when the solver sets the household onto the borrowing constraint.
This is problematic when the household is predicted to be constrained, but should not be.
In such cases, $\frac{\hat{b}^{h+1}_{t+1} - \underline{b}}{\hat{c}_t^h} = 0$, while $\frac{(u')^{-1}\biground{\frac{\beta \Eind{t}{u'(\hat{c}_{t+1}^{h+1})(1 + \zeta^b (\hat{b}_{t+2}^{h+2} - \hat{b}_{t+1}^{h+1}))}}{p_t + \zeta^b (\hat{b}_{t+1}^{h+1} - \hat{b}_{t}^{h})}}}{\hat{c}_t^h} -1 < 0$.
This would correctly lead to a positive term in the loss function, $\biground{\text{err}^{c, h}_\rho(\x_t)}^2>0$, but to no feedback to the neural network parameters to increase the prediction $\tilde{b}^h$.
To restore this feedback, we add the square of the following term to the loss function
\begin{align}
\text{err}_\rho^{h, \text{false binding}}(\x_t) &:= \frac{1}{1+\tilde{b}^{h+1}_{t+1}} \times e^{-\frac{\left(\hat{b}^{h+1}_{t+1} - \underline{b}\right)^{2}}{10^{-5}}} \times \max \left\{-\frac{(u')^{-1}\biground{\frac{\beta \Eind{t}{u'(\hat{c}_{t+1}^{h+1})(1 + \zeta^b (\hat{b}_{t+2}^{h+2} - \hat{b}_{t+1}^{h+1}))}}{\hat{p}^b_t + \zeta^b (\hat{b}_{t+1}^{h+1} - \hat{b}_{t}^{h})}}}{\hat{c}_t^h}+ 1,0\right\}. \label{eq:liftoff}
\end{align}
The intuition behind these terms is as follows.
The last term, $\max \left\{-\frac{(u')^{-1}\biground{\frac{\beta \Eind{t}{u'(\hat{c}_{t+1}^{h+1})(1 + \zeta^b (\hat{b}_{t+2}^{h+2} - \hat{b}_{t+1}^{h+1}))}}{p_t + \zeta^b (\hat{b}_{t+1}^{h+1} - \hat{b}_{t}^{h})}}}{\hat{c}_t^h}+1,0\right\}$, is always non-negative and only different from zero, when the household is currently saving too little, in which case it should not be on the borrowing constraint.
The second term, $e^{-\frac{\left(\hat{b}^{h+1}_{t+1} - \underline{b}\right)^{2}}{10^{-5}}}$, is a differentiable approximation to a function, which is always zero, except when $\hat{b}^{h+1}_{t+1} - \underline{b}$, in which case it is equal to one. Hence, it is always non-negative and only positive if the household is predicted to be constrained.
Taken together, the last two terms in equation \eqref{eq:liftoff} ensure that the term is only positive if the household is falsely predicted to be constrained. Finally, the first term, $\frac{1}{1+\tilde{b}^{h+1}_{t+1}}$, ensures that the overall error term is reduced by increasing $\tilde{b}^{h+1}_{t+1}$ and is thus restoring a pass through from the loss function to the neural network parameters governing the prediction for the household's policy.

\paragraph{Integration}
An evaluation of the loss function implied by the model requires the computation of conditional expectations in the Euler equations, which characterize the households' behavior. Despite a large number of state variables, this economy features only one source of uncertainty: a stationary AR(1) process with Gaussian innovations. This structure allows us to approximate conditional expectations using standard Gauss-Hermite quadrature \cite[for more details on numerical integration in the context of economic modeling, see][]{judd1998numerical}. We denote the number of quadrature nodes with $N^{\text{integration}}$ and choose $N^{\text{integration}} = 8$.

\subsubsection{Training}\label{sec:single_training}
 To parameterize equilibrium objects of the economy (policies and prices), we use a densely connected feed-forward neural network with two hidden layers. The dimensionality of its input is $N^{\text{input}} = 21$ since the state vector includes both, the aggregate productivity shock, and the asset distribution across all twenty age groups in the economy. For the two hidden layers, we choose $N^{\text{hidden 1}} = N^{\text{hidden 2}} = 400$ neurons with relu activation functions.\footnote{The relu activation function is given by $\text{relu}(x):=\max\{0, x\}$. Relu is short for the rectified linear unit.} Finally, the network output is a vector of $N^{\text{output}}=41$ elements consisting of bond savings, housing services consumption policy functions, and two price functions for the price of the bond and the price for renting housing services. Outputs corresponding to prices are transformed using the softplus activation in order to ensure that the predicted prices are always positive.\footnote{The softplus activation function is given by $\text{softplus}(x) := \log(1+\exp(x)$.}

To train the network to approximate equilibrium price and policy functions, we use the training procedure described in section \ref{sec:learn_alg}. To obtain training states, we simulate $N^{\text{trajectories}} = 8192$ parallel state trajectories for $N^{\text{episodes}} = 3584$ episodes. For each episode (\emph{i.e.} each simulation step), we train the neural network for $N^{\text{epochs}} = 10$ epochs. In each epoch, we split the simulated dataset of $N^{\text{trajectories}}$ states into batches of $N^{\text{minibatch}} = 128$ states. Each batch is used to compute one stochastic gradient descent update of the neural network weights. To calculate those updates, we use the Adam optimizer \citep[][]{kingma2014adam}, a version of the stochastic gradient descent algorithm, with a learning rate of $\alpha^{\text{learn}} = 1\times 10^{-5}$.\footnote{To improve robustness of our learning algorithm, we transformed loss function gradients using the \texttt{$\text{zero} \_ \text{nans}$} transformation before using them as an input for the Adam optimizer. Specifically, we use \texttt{adam} and \texttt{zero\_nans} provided by the \texttt{optax} library:  \url{https://optax.readthedocs.io/en/latest/}. The \texttt{zero\_nans} transformation replaces \texttt{NaN},when they occur, with 0, avoiding a failure of the training. Furthermore, we found it useful to re-start internal state of Adam optimizer to zero at the beginning of training episode.}

\begin{table}[tb!]
\begin{center}
\begin{tabular}{ cccccccc}
\toprule
Parameters & $N^{\text{input}}$ & \makecell{$N^{\text{hidden 1}}$\\ Activations} & \makecell{$N^{\text{hidden 2}}$\\ Activations}  & \makecell{$N^{\text{output}}$\\ Activations}\\
\midrule
Values & 21 & \makecell{400\\ relu} & \makecell{400 \\ relu} & \makecell{41\\ see text} \\
\bottomrule
\end{tabular}
\caption{Network Architecture chosen for the single asset model.\label{tab:network_simple}}
\end{center}
\end{table}

\begin{table}[tb!]
\begin{center}
\begin{tabular}{ ccccccccc}
\toprule
Parameters & $N^{\text{episodes}}$ & $N^{\text{trajectories}}$ & $N^{\text{epochs}}$& $N^{\text{minibatch}}$ & $N^{\text{integration}}$ & $\alpha^{\text{learn}}$ \\
\midrule
Values & 3584 & 8192 & 10 & 128 & 8 & $10^{-5}$\\
\bottomrule
\end{tabular}
\caption{Hyperparameters chosen for the single asset model.\label{tab:hyperpar_simple}}
\end{center}
\end{table}

\subsubsection{Accuracy}
\begin{figure}
\centering
\includegraphics[width = 0.45\textwidth]{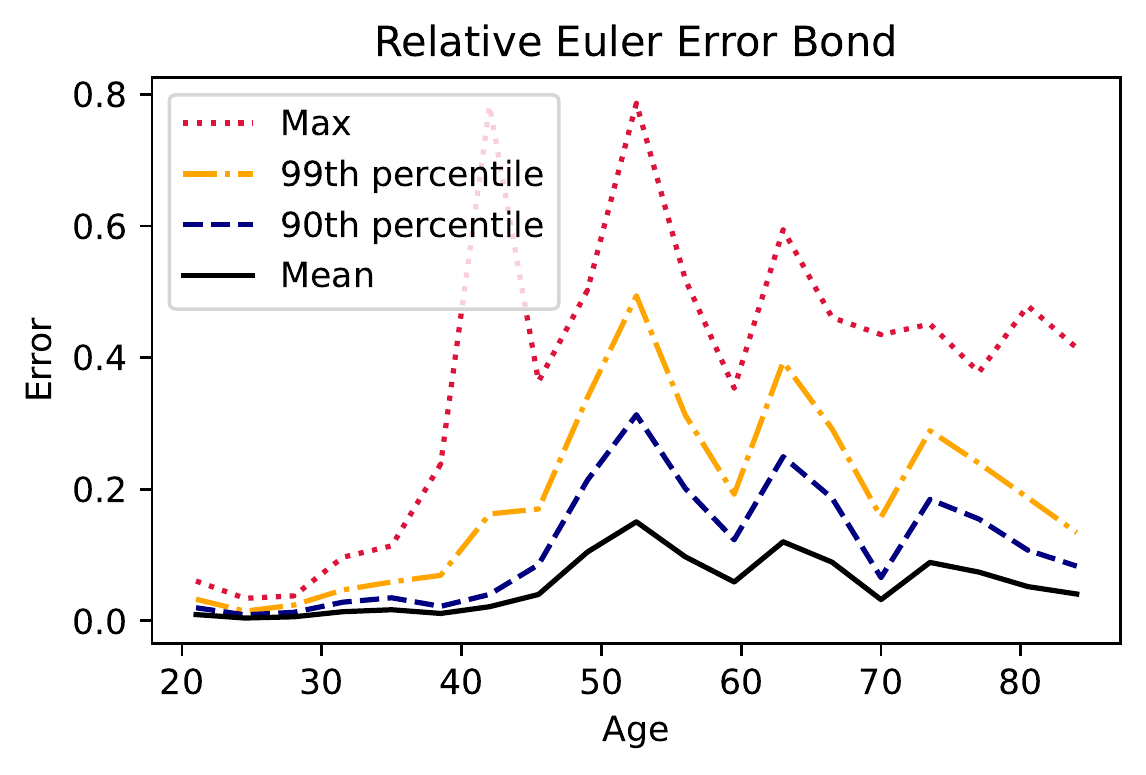}
\includegraphics[width = 0.45\textwidth]{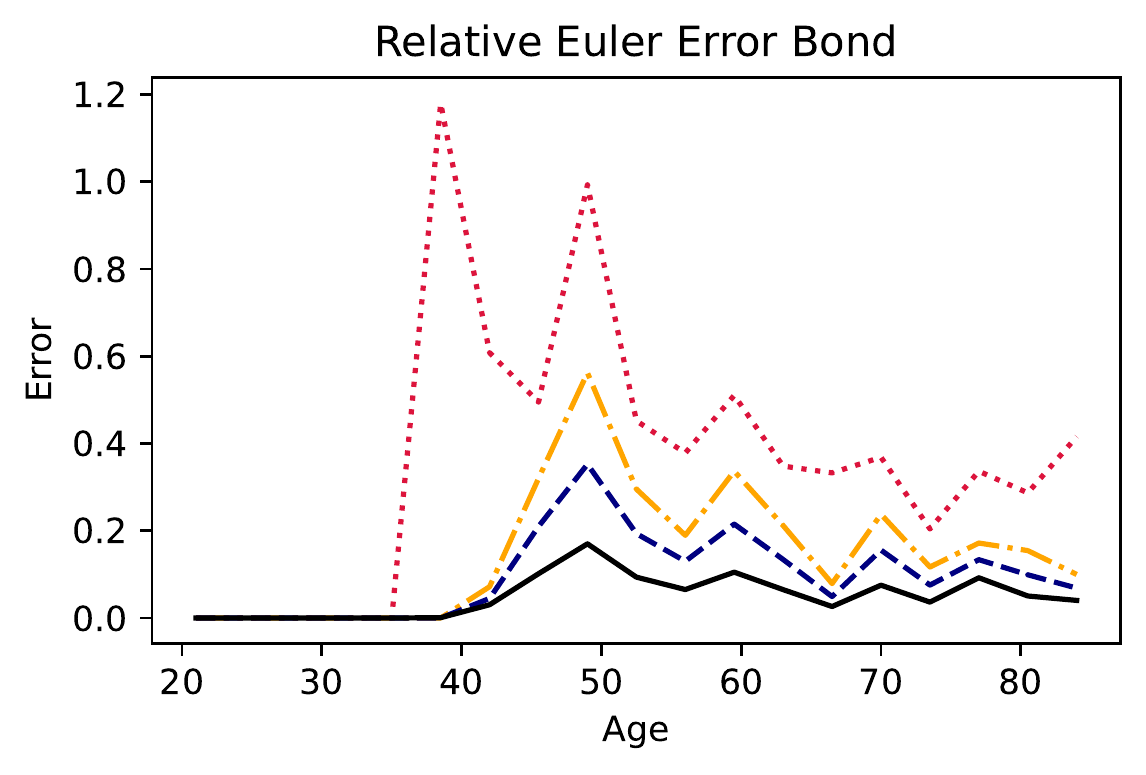}\\
\includegraphics[width = 0.45\textwidth]{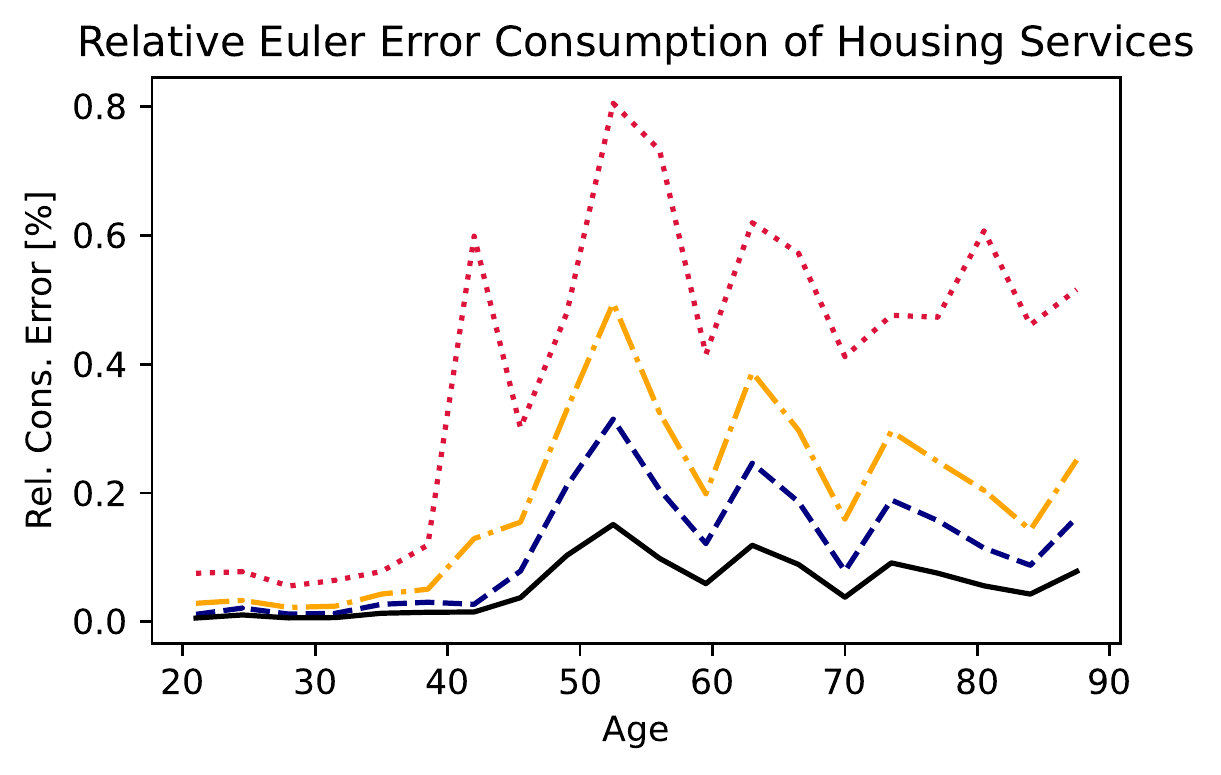}
\includegraphics[width = 0.45\textwidth]{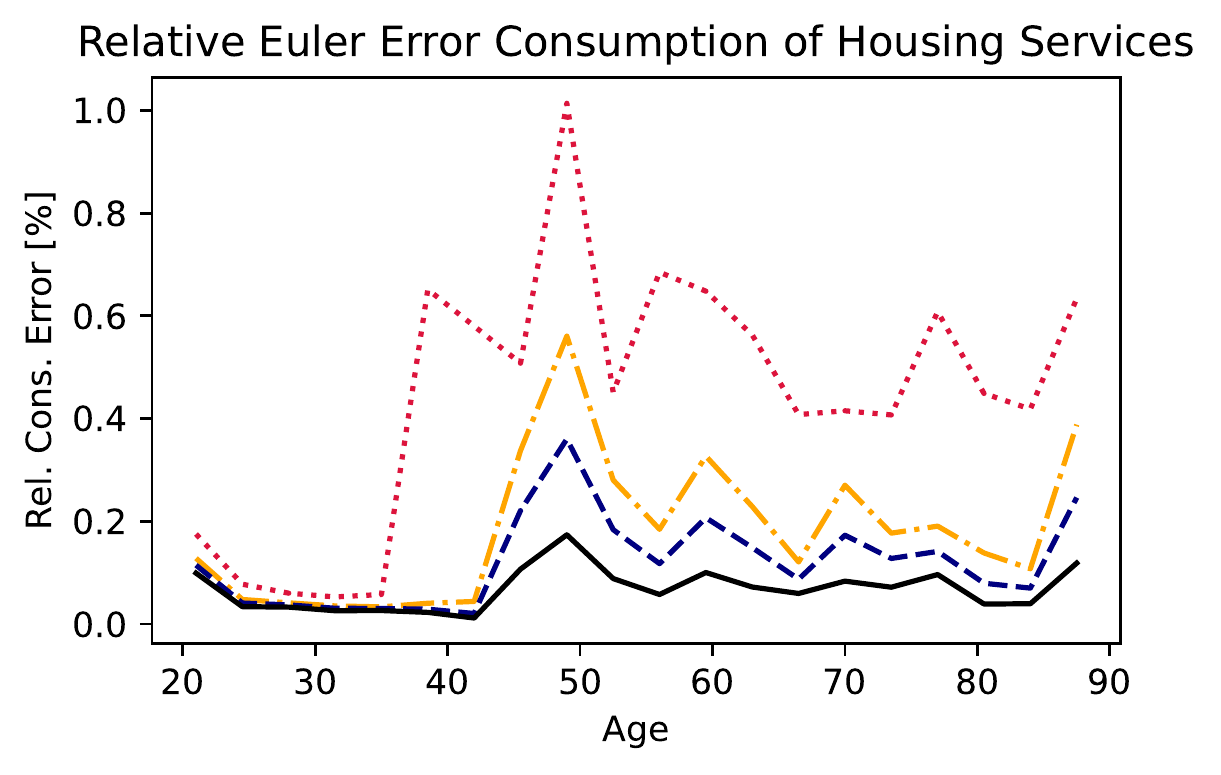}
\caption{\label{fig:simple_errors}Statistics on the violation of the equilibrium conditions by age group on a simulated path of $8192$ states by age group. The left panel shows the statistics for the simple market clearing architecture and the right panel shows the results for the solver-based architecture, which additionally guarantees that the borrowing constraints are always satisfied. 
The top row shows the errors in the KKT conditions for the bond and the bottom row shows  the errors in the Euler equation for the renting of housing services.
The different lines, from the lowest to the highest, show the minimum, the 10th percentile, the average, the 90th percentile and the maximum.}
\end{figure}

To evaluate the quality of our solution, we evaluate statistics of the errors in the equilibrium conditions along simulated paths of the economy.
We simulate a total of $2^{13} = 8192$ new states without retraining the neural network.
Figure \ref{fig:simple_errors} reports statistics on the remaining errors in the KKT conditions for each age group, expressed as relative consumption errors.
As the figure shows, the models are solved precisely, with the 99th percentile of errors in the optimality conditions well below 1\% for each age group and both policies.
The left panel in figure \ref{fig:simple_errors} shows the simple market clearing architecture described in section \ref{sec:simple}, the right panel shows the solver-based architecture, which additionally guarantees that the borrowing constraints are always satisfied.
Indeed we can see that the equilibrium errors are fulfilled exactly for the young agents, for which the borrowing constraint is always binding.

\section{Homotopy Algorithm for Portfolio Choice}\label{sec:homotopy_loop}
While deep learning based solution methods can handle high-dimensional state spaces comparatively well, portfolio choice problems pose a challenge for this approach. Portfolio choice is pinned down only at low levels of errors in the households' optimality conditions. In the beginning of training the errors are still high and it is therefore hard to learn the portfolio choice correctly.
As a consequence, the training process is prone to get stuck at spurious solutions featuring a bang-bang type of portfolio allocations. In general, those oscillations are difficult to overcome. 

In simulation-based solution algorithms, this issue is amplified, since the intermediate ``solution'' is used to simulate the dynamics of the economy to produce new state values, on which the network is subsequently trained. Bad policies might hence generate economically non-sensible states, which are far from the domain on which the neural network was trained previously. Those oscillations can hence create a feedback loop, which might hamper the convergence of the whole algorithm. 

To ameliorate the propensity of portfolio choice models to generate zig-zag oscillations, we propose a homotopy procedure.
Homotopy methods \citep[see][]{schmedders_1998} are a general class of methods.
The idea is to start by solving a simple problem and then slowly transform the problem into the harder problem of interest.
If the transformations of the problem are gradual enough, the previously obtained solution to the nearby problem provides a good starting point to solve the slightly harder, transformed problem.
In order to leverage this concept for solving portfolio choice models, we start by solving a single-asset economy, which is then gradually enriched with new assets. By starting with solving a simple problem up to a high degree of precision, and then creating a sequence of perturbed problems, where every two consecutive problems are very similar to each other, we are able to guide the solution procedure and avoid spurious zig-zag solutions, which may otherwise prevent the solution procedure from converging to a high-quality solution. 

For a practical implementation of the homotopy procedure, we utilize a natural nesting structure present in portfolio choice models. A multi-asset economy can be transformed into a single-asset one by setting the net supply of all but one asset to zero and imposing a strict no short-sale constraint. Given a zero net supply and a strict no short-sale constraint, equilibrium allocations of that asset are uniformly zero. Moreover, the loss function used to train the network can be parameterized with weights on Euler equations of different assets. By setting the weights on Euler equations for some assets to zero and and imposing zero net supply coupled with a strict no short-sale constraint, the loss function constructed to represent the multiple-asset economy characterizes the single-asset economy. In addition to parameters governing asset supply and the weights in the loss function, we also employ a mask parameter that multiplies all the asset policies. We initially set the mask to zero for all assets, which are not yet included into the economy.
The mask parameter is increased to small positive values when the corresponding asset is added to the economy.

This nesting structure allows us to implement our homotopy procedure using a simple loop that implements the deep learning procedure for a particular parameterization of the economy and the loss function. When adding the N-th asset, we first use the $N-1$ asset economy and the N-th asset Euler equations to construct a price function for the N-th asset. Next, we introduce a small amount of the N-th asset and take the policy and the price functions constructed in the previous step as an initial guess. Then we run the deep learning procedure to solve for an equilibrium of this perturbed economy.

Our homotopy procedure can be summarized in the following steps:

\begin{itemize}
    \item Parameterize all price and policy functions using a neural network equipped with market clearing layers. This network takes as an input a full state vector of the multi-asset economy and outputs all prices and portfolio allocations. Initialize its parameters.

    \item For all but one asset set the net supply to zero, and impose strict no short-sale constraints on the remaining assets. Set the weights in the loss function on Euler equations associated with the assets in zero net supply to zero. Set the mask parameter for the corresponding policy functions to zero.
    
    \item Select initial state values for the training procedure.
    
    \item Starting from initial weights and state values, run the training procedure to solve for an equilibrium of the single asset economy.

    \item Set the weights in the loss function on the Euler equations associated with the second asset to its full value (e.g. 1). Keep the zero net supply and the no short-sale constraint.
    
    \item Use the Euler equations associated with the second asset to solve for its price function. Since the weight on this Euler equations is non-zero, the network will be able to learn a price function implied by zero allocations of that asset. Recall that this second asset is in zero net supply, and agents are bound by strict no short-sale when trading it.

    \item Set the mask parameter for the second asset to a small, but non-zero value (\emph{e.g.} 0.01). Now the network can predict non-zero allocations of the second asset. The choice of a small mask value helps to avoid instability associated with a potentially large and discontinuous jump. 
    
    \item Set the net supply of the second asset at a small, but non-zero value. Use network weights and state values from the previous step as an initial guess for solving this economy.
    
    \item Increase the net supply of the second asset by another step. Re-solve the economy starting from weights and states from the previous step. Repeat this procedure until the supply target for the second asset is reached.
    \item If required, perform the same procedure with no-short sale constraint (\emph{i.e.} relax it slowly).
    
    \item Set the weight in the loss function on the Euler equations associated with the third asset to its full value (\emph{e.g.} 1).
    
    \item Since the weight on the Euler equations is non-zero, the network will be able to learn a price function implied by zero allocations of that asset. Recall that this third asset is in zero net supply, and agents are bound by strict no short-sale when trading it.

    \item Set the mask parameter for the third asset to a small, but non-zero value (\emph{e.g.} 0.01). Now the network can predict non-zero allocations of the third asset. The choice of a small mask value helps to avoid instability associated with a potentially large and discontinuous jump. 
    
    \item Set the net supply of the third asset at some small, but non-zero level. Solve this economy starting from network weights from the previous step.
    
    \item Increase the net supply of the third asset by another step. Re-solve the economy starting from weights and states from the previous step. Repeat this procedure until the supply target for the third asset is reached.

    \item If required, perform the same procedure with no-short sale constraint (\emph{i.e.} relax it slowly).

    \item In case of adding the fourth and further asset, perform the analogous procedure.
\end{itemize}

\subsection{Example with Three Assets}
\subsubsection{Model\label{sec:multiasset_model}}
To illustrate the homotopy method, together with our market-clearing neural network architecture, we enrich the overlapping generations model introduced in section \ref{sec:simple_model} with two additional assets.
Next to a one-period risk-free bond, households can also trade claims to a Lucas tree as well as claims to the stream of rent payments for housing services. We think of the former as a stock and of the latter as house ownership.
The households hence have to allocate their total savings between three different assets.
This portfolio choice problem is challenging for deep learning based solution methods, rendering the model a suitable testing ground for our homotopy method.
The remaining components, are the same as in the model we previously described in \ref{sec:simple_model}.
\paragraph{Asset structure}
As before, households can purchase a risk-free one-period bond in net-supply $B$ at equilibrium prices $p^b_t$.
Additionally, households can trade claims to a Lucas tree at equilibrium prices $p^s_t$, which in net-supply $S$.\footnote{The abbreviation $S$ stands for stock.}
Every period, the Lucas Tree pays an exogenous amount of dividends $d_t = d z_t$, which is perfectly correlated with aggregate labor income.
Next to deriving utility from  consumption, household derive age-specific utility from consuming housing services, which they rent every period at equilibrium rent $p^{r}_t$.
Finally households can trade claims to the aggregate housing stock at equilibrium price $p^{o}_t$, which is in net-supply $H^o$.
The total housing stock in the economy, which can be rented by household is the sum of the locally owned housing supply $H^o$, which are the houses owned by households, and an housing supply $H^{ex}$, with owners outside the model.
In this model a claim to the housing stock is a claim to the streams of rents. Owning a house is hence similar to owning a Lucas tree, where the dividends are replaced by the rent payments the house owners receive.
None of the assets can be sold short, and all of the assets are traded subject to asset-specific quadratic adjustment costs $p^x_t \zeta^x \frac{1}{2} (x^{h+1}_{t+1} - x^h_t)^2$, for $x\in\{b, s, o\}$.
We model the bond as the most liquid asset and house ownership as the least liquid asset, \emph{i.e.} $\zeta^h > \zeta^s  > \zeta^b$.
The Bellman equation corresponding to the households' problem is given by
\begin{align}
V^h(z_t, b_t^h, s_t^h, h^{o, h}_{t}) &= \max_{h^h_{r,t},b_{t+1}^{h+1}, s_{t+1}^{h+1}, h^{o, h+1}_{t+1}}\left\{u(c_t^h) + \psi^h v(h^{r, h}_{t}) + \beta \E{V^{h+1}(z_{t+1}, b_{t+1}^{h+1}, s_{t+1}^{h+1}, h^{o, h+1}_{t+1})}\right\}\\
c_t^h &= z_t y^h + b_t^h + s_t^h (p_t^s + d_t^s) + h^{o, h}_{t} (p_t^{o} + p_t^r) \nonumber \\
&- b_{t+1}^{h+1} p_t^b - s_{t+1}^{h+1} p_{t+1}^s - h^{o, h+1}_{t+1} p_t^{o} - h^{r, h}_{t}p_t^{r} \nonumber \\
&-p_t^b\zeta^b \frac{1}{2} \biground{b_{t+1}^{h+1} - b_{t}^{h}}^2-p_t^s\zeta^s \frac{1}{2} \biground{s_{t+1}^{h+1} - s_{t}^{h}}^2-p_t^o\zeta^h \frac{1}{2} \biground{h_{t+1}^{o, h+1} - h_{t}^{o, h}}^2\\
\text{subject to}:& \nonumber \\
0 &\leq b_{t+1}^{h+1} - \underline{b}  \\
0 &\leq s_{t+1}^{h+1} - \underline{s}  \\
0 &\leq h_{t+1}^{o, h+1} - \underline{h}^o 
\end{align}
The corresponding optimality conditions, which, together with market clearing, characterize the functional rational expectations equilibrium, are given in Appendix \ref{app:hh_prob}. We provide details on parameter choices in Appendix \ref{app:calib_het}.

\paragraph{Heterogeneity in risk aversion}
We continue to model an overlapping generations economy, where households live deterministically for $H$ periods and receive an age specific share of aggregate labor income.
Additionally, we now assume that in each cohort, there are two types of households of equal mass.
One type has a lower risk aversion, the other type as a higher risk aversion.
\paragraph{State of the economy}
The state of the economy is given by the exogenous shock, as well as the distribution of assets accross households and risk-aversion types.
Let the superscript $^1$ denote the low risk aversion households and superscript $^2$ denote the high risk aversion households.
The state of the economy is given by
\begin{align}
\x_t = [z_t, \underbrace{b_t^{1, 1}, \dots, b_t^{H, 1}, b_t^{1, 2}, \dots, b_t^{H, 2}}_{\text{bond holdings}}, \underbrace{s_t^{1, 1}, \dots, s_t^{H, 1}, s_t^{1, 2}, \dots, s_t^{H, 2}}_{\text{stock holdings}}, \underbrace{h_t^{o, 1, 1}, \dots, h_t^{o, H, 1}, h_t^{o, 1, 2}, \dots, h_t^{o, H, 2}}_{\text{housing ownership}}]
\end{align}
The state space is hence $1 + 2\times 3 \times (H- 1)$-dimensional.\footnote{The $(H - 1)$ stems from the fact that agents are born without assets and their asset holding is hence always constant and does not need to be included in the state.}
All policy and price functions are a function of the whole state, rendering this model a formidable testing ground for our algorithm.
\paragraph{Equilibrium functions to approximate}
We need to approximate $2\times (H- 1) \times 4 + 2$ household policies: for each risk aversion type and each age group, we need the bond policy, the stock policy, the renting policy as well as the house ownership policy.
While we only need to approximate $H - 1$ policies per asset and risk aversion type, we need to approximate $H$ policies for the intra-temporal choice of renting housing services.
Furthermore, we need to approximate four equilibrium price functions: for the bond price, the tree price, the price for renting, and the price for house owner ship.

We approximate the mapping from the $1 + 2\times 3 \times (H- 1)$ dimensional state to the $2\times (H- 1) \times 4 + 2+ 4$ endogenous variables by a deep neural network.
\begin{align}
\nn &: \mathbb{R}^{1 + 2\times 3 \times H} \rightarrow \mathbb{R}^{2\times (H- 1) \times 4 + 2 + 4}\nonumber\\
\nn(\x) = [&\tilde{b}^{2, 1}_{t+1}, \dots, \tilde{b}^{H, 1}_{t+1}, \tilde{b}^{2, 2}_{t+1}, \dots, \tilde{b}^{H, 2}_{t+1}, \nonumber\\
&\tilde{s}^{2, 1}_{t+1}, \dots, \tilde{s}^{H, 1}_{t+1}, \tilde{s}^{2, 2}_{t+1}, \dots, \tilde{s}^{H, 2}_{t+1},\nonumber\\
&\tilde{h}^{o, 2, 1}_{t+1}, \dots, \tilde{h}^{o, H, 1}_{t+1}, \tilde{h}^{o, 2, 2}_{t+1}, \dots, \tilde{h}^{o, H, 2}_{t+1},\nonumber\\
&\tilde{h}^{r, 1, 1}_{t}, \dots, \tilde{h}^{r, H, 1}_{t}, \tilde{h}^{r, 2, 1}_{t}, \dots, \tilde{h}^{r, H, 2}_{t},\nonumber\\
&\hat{p}^b_t, \hat{p}^s_t, \hat{p}^{o}_t, \hat{p}^{r}_t]
\end{align}
We use the simple adjustment described in section \ref{sec:simple} to ensure that market clearing conditions are always satisfied and use our homotopy algorithm to solve the model step-wise by slowly increasing the supply and number of assets.
\paragraph{Loss function}
Since our neural network architecture already enforces market clearing, the only remaining equilibrium conditions are the households' optimality conditions for each of the four choices.
Analogously to equation \eqref{eq:hh_fb_bond}, we use the Fischer Burmeister equation to summarize each set of KKT conditions into a single equation.
Let $\left\{\{\text{err}_\rho^{b, i, h}(\x_t), \text{err}_\rho^{s, i, h}(\x_t),  \text{err}_\rho^{{o}, i, h}(\x_t)\}_{h = 1}^{H - 1}, \{\text{err}_\rho^{{h^r}, i, h}(\x_t)\}_{h = 1}^H\right\}_{i = 1}^2$ denote the corresponding errors in the equilibrium conditions.
Our loss function is given by
\begin{align}
l_{\mathbf{\rho}}(\mathcal{D}) &:= \frac{1}{2}\left( \right.\nonumber\\
&w_b \frac{1}{|\mathcal{D}|}\frac{1}{H - 1}\left(\sum_{\x \in \mathcal{D}} \sum_{h = 1}^{H - 1} \biground{\text{err}_\rho^{b, 1, h}(\x_t)}^2 + \biground{\text{err}_\rho^{b, 2, h}(\x_t)}^2\right)\nonumber\\
&+w_s \frac{1}{|\mathcal{D}|}\frac{1}{H - 1}\left(\sum_{\x \in \mathcal{D}} \sum_{h = 1}^{H - 1} \biground{\text{err}_\rho^{s, 1, h}(\x_t)}^2 + \biground{\text{err}_\rho^{s, 2, h}(\x_t)}^2\right) \nonumber\\
&+w_{o} \frac{1}{|\mathcal{D}|}\frac{1}{H - 1}\left(\sum_{\x \in \mathcal{D}} \sum_{h = 1}^{H - 1} \biground{\text{err}_\rho^{{o}, 1, h}(\x_t)}^2 + \biground{\text{err}_\rho^{{o}, 2, h}(\x_t)}^2\right) \nonumber\\
&+w_{h^r} \frac{1}{|\mathcal{D}|}\frac{1}{H}\left(\sum_{\x \in \mathcal{D}} \sum_{h = 1}^{H} \biground{\text{err}_\rho^{{r}, 1, h}(\x_t)}^2 + \biground{\text{err}_\rho^{{r}, 2, h}(\x_t)}^2\right) \nonumber\\
)
\end{align}
The weights $w_x$ allow to vary the weight on the equilibrium conditions.
The loss function hence nests a single and two-asset model, by allowing the corresponding weight in the loss function to be set to zero, while also ensuring that the associated policies are equal to zero.

\subsubsection{Training}\label{sec:het_training}
To parameterize equilibrium objects of the economy (\emph{i.e.}  policies and prices), we use a densely connected feed-forward neural network with two hidden layers. The dimensionality of its input is $N^{\text{input}} = 161$ since the state vector includes aggregate productivity shock and asset distribution across all twenty age groups for both types in the economy, and the distribution of dividend income, which we choose as an auxiliary state variable.\footnote{The minimal state vector includes $121$ state variables, that includes distribution of three assets across $20$ age groups for each of two risk aversion groups and the aggregate shock. Following \cite{azinovic2022deep}, we augmented the input of the neural network by a vector of auxiliary statistics. In our case, we choose a distribution of dividend income as an auxiliary statistic.} For the two hidden layers, we choose $N^{\text{hidden 1}} = N^{\text{hidden 2}} = 400$ neurons with relu activation functions. Finally, the network output is a vector of $N^{\text{output}}=158$ elements consisting of the policies for the three assets, the policies for renting housing services, and four price functions for the price of the bond, the price of stocks, the price for house ownership, and the price for renting housing services.\footnote{For each asset category, we approximate the $19$ savings function, exploiting the fact that the oldest age group does not save in our model. Furthermore, we approximate the $20$ housing service consumption function, since housing services are demanded by all age groups.} Outputs corresponding to prices are transformed using the softplus activation in order to ensure that the predicted prices are always positive.

To ameliorate numerical instabilities associated with portfolio choice, we employ the homotopy algorithm described at the beginning of section \ref{sec:homotopy_loop}. Specifically, we firstly solve a bond-only economy. To do so, we set the $S^{\text{initial}} = 0 $, $H^{\text{o,initial}} = 0$, and $H^{\text{ex,initial}} = 1$ and run the training procedure described in section \ref{sec:learn_alg}.\footnote{We use the same training hyperparameters as in subsequent homotopy steps, with exception of $N^{\text{episodes}}$, which we set to $512$, in order to obtain a high-precision solution to start from. The value for the remaining hyperparameter are provided in table \ref{tab:hyperpar_hetrra}.} Then, we approximate the stock pricing function implied by the consumption dynamics of the bond-only economy.\footnote{In this step, we use the same training hyperparameters as in subsequent homotopy steps. See table \ref{tab:hyperpar_hetrra}.} Using the price and policy functions of the bond-only economy as an initial guess, we compute equilibrium of the two asset economy featuring a small supply of stocks. Starting from this solution, we gradually increase stock supply and retrain the neural network after each increase. We do so until we reach the stock supply target. In total, we perform $S^{\text{steps}} = 10$ homotopy steps to increase the stock supply from $S^{\text{initial}} = 0$ to $S^{\text{final}} = 1$. We proceed analogously for the case of housing ownership. In that case, we perform $H^{\text{steps}} = 20$ housing supply steps. We simultaneously increase the supply of private housing $H^{\text{o}}$ from $H^{\text{o, initial}} = 0$ to $H^{\text{o, final}} = 1$ and decrease the external housing supply $H^{\text{ex}}$ from $H^{\text{ex, initial}} = 1$ to $H^{\text{ex, final}} = 0$, such that the total housing supply $H^o + H^{ex}$ remains constant and equal to one.

In each step of the homotopy algorithm we continue to train the neural network to approximate the equilibrium functions for the economy with the corresponding values for the aggregate supply of assets in the economy, \emph{i.e.} for the intermediate values for $(B, S, H^o, H^{ex})$. Again, we use the learning procedure described in section \ref{sec:learn_alg}. To obtain training states, we simulate $N^{\text{trajectories}} = 8192$ parallel state trajectories for $N^{\text{episodes}} = 256$ episodes.\footnote{In each homotopy step, we obtain the initial set of states and initial network weights from the terminal value of previous homotopy step.} For each episode (\emph{i.e.} each simulation step), we train the neural network for $N^{\text{epochs}} = 10$ epochs with a learning rate of $\alpha^{\text{learn}} = 1\times 10^{-6}$.\footnote{In each epoch, we split the simulated dataset of $N^{\text{trajectories}}$ states into batches of $N^{\text{minibatch}} = 128$ states. Each batch is used to compute one stochastic gradient descent update of the neural network weights. As for the single asset model, we transformed loss function gradients using the \texttt{$\text{zero} \_ \text{nans}$} transformation before using them as an input for the Adam optimizer and re-start internal state of Adam optimizer to zero at the beginning of training episode.}
\begin{table}[tb!]
\begin{center}
\begin{tabular}{ cccccccc}
\toprule
Parameters & $N^{\text{input}}$ & \makecell{$N^{\text{hidden 1}}$\\ Activations} & \makecell{$N^{\text{hidden 2}}$\\ Activations}  & \makecell{$N^{\text{output}}$\\ Activations}\\
\midrule
Values & 161 & \makecell{400\\ relu} & \makecell{400 \\ relu} & \makecell{158\\ see text} \\
\bottomrule
\end{tabular}
\caption{Network Architecture chosen for the three asset model with risk aversion heterogeneity.\label{tab:network_threeasset}}
\end{center}
\end{table}
\begin{table}[tb!]
\begin{center}
\begin{tabular}{ccccccccc}
\toprule
Parameters & $N^{\text{episodes}}$ & $N^{\text{trajectories}}$ & $N^{\text{epochs}}$ & $N^{\text{minibatch}}$ & $N^{\text{integration}}$ & $\alpha^{\text{learn}}$ \\
\midrule
Values & 256 & 8192 & 10 & 128 & 8 & $10^{-6}$\\
\bottomrule
\end{tabular}
\caption{Hyperparameters for training steps within homotopy loop.\label{tab:hyperpar_hetrra}}
\end{center}
\end{table}
\begin{table}[tb!]
\begin{center}
\begin{tabular}{ccccccccccc}
\toprule
Parameters & $S^{\text{initial}}$ & $S^{\text{final}}$ & $H^{o, \text{initial}}$ & $H^{o, \text{final}}$ &  $H^{ex, \text{initial}}$ & $H^{ex, \text{final}}$ & $S^{\text{steps}}$ & $H^{\text{steps}}$ \\
\midrule
Values & 0 & 1 & 0 & 1 & 1 & 0 & 10 & 20\\
\bottomrule
\end{tabular}
\caption{Homotopy loop hyperparameters \label{tab:homotopy_hetrra}}
\end{center}
\end{table}

\begin{figure}
\centering
\includegraphics[width = 0.32\textwidth]{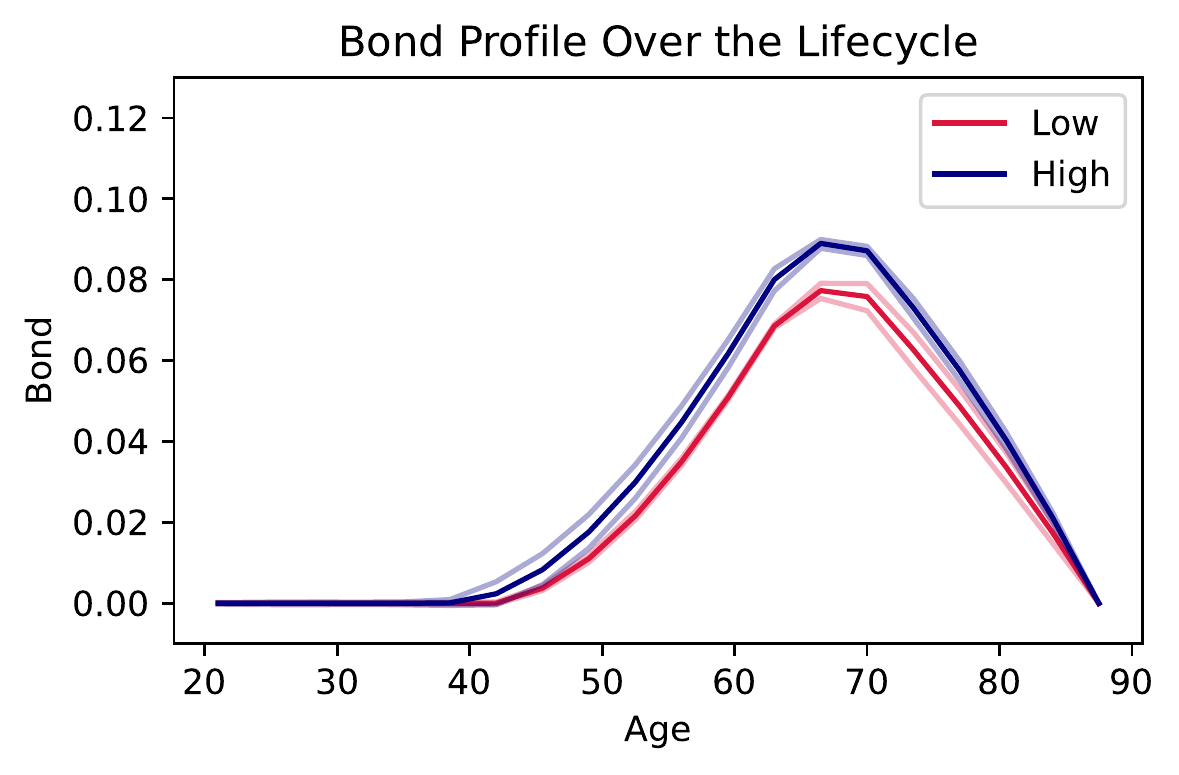}
\includegraphics[width = 0.32\textwidth]{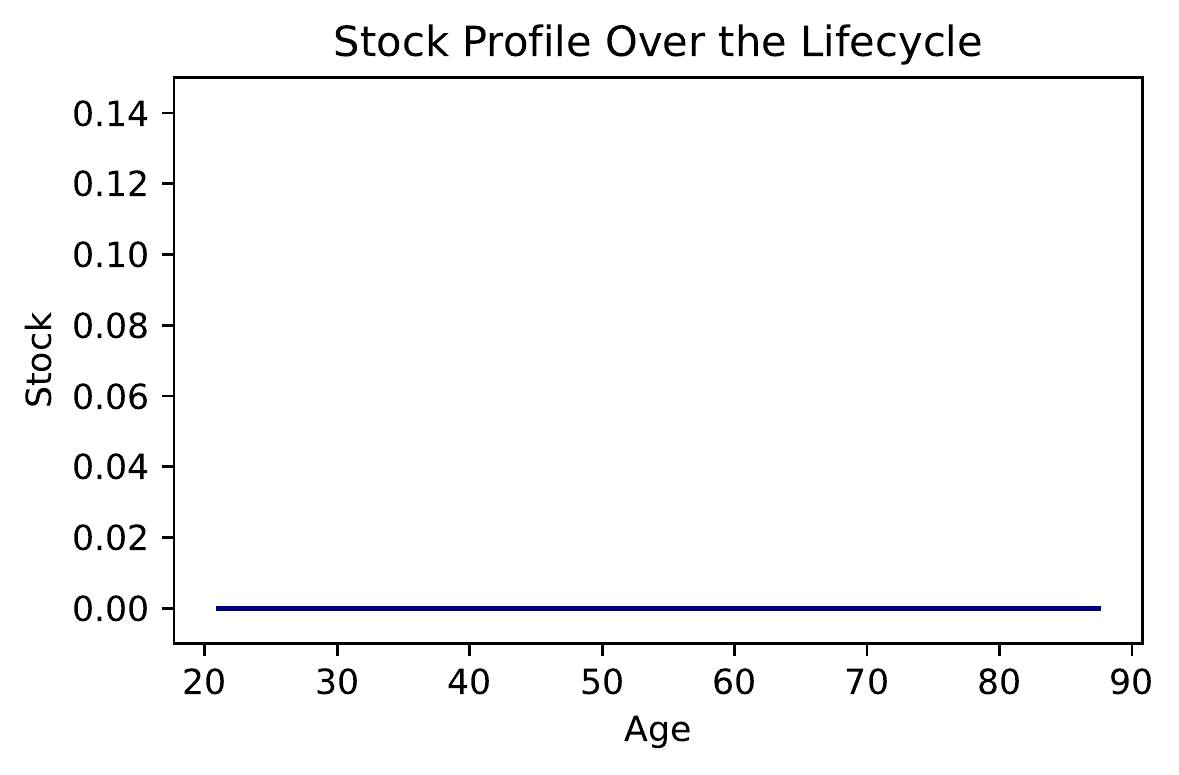}
\includegraphics[width = 0.32\textwidth]{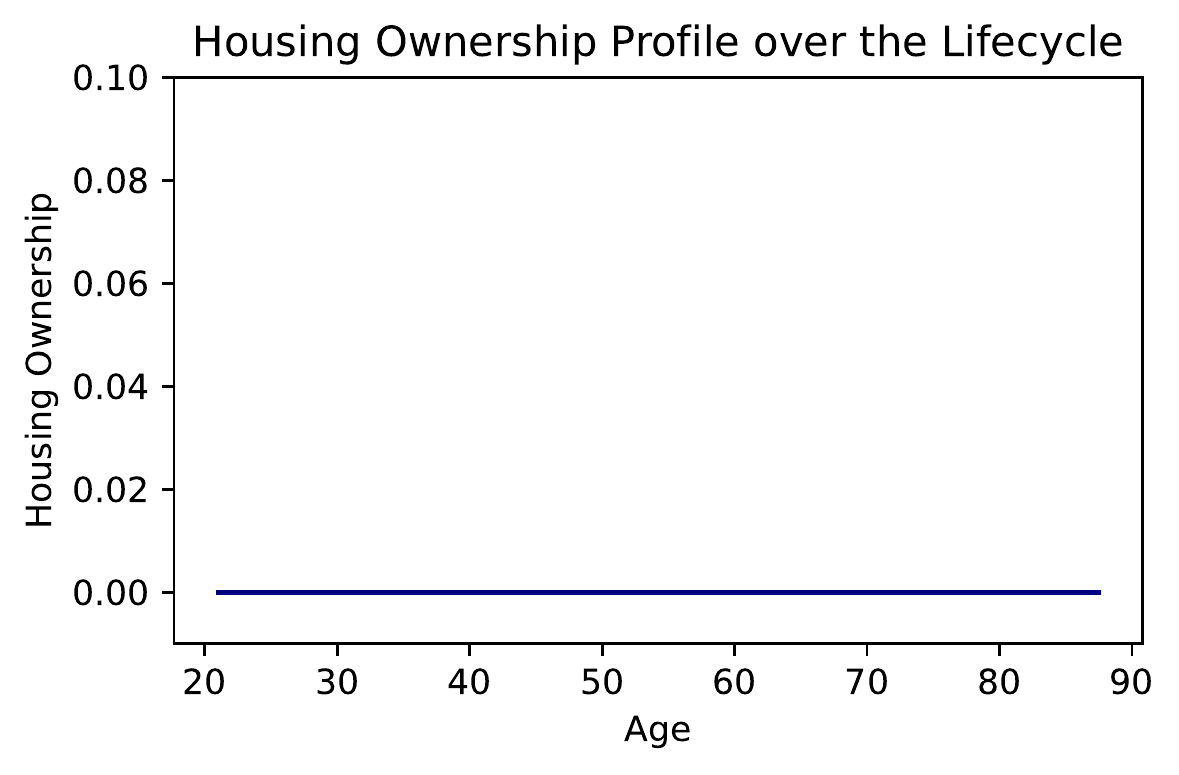}\\
\includegraphics[width = 0.32\textwidth]{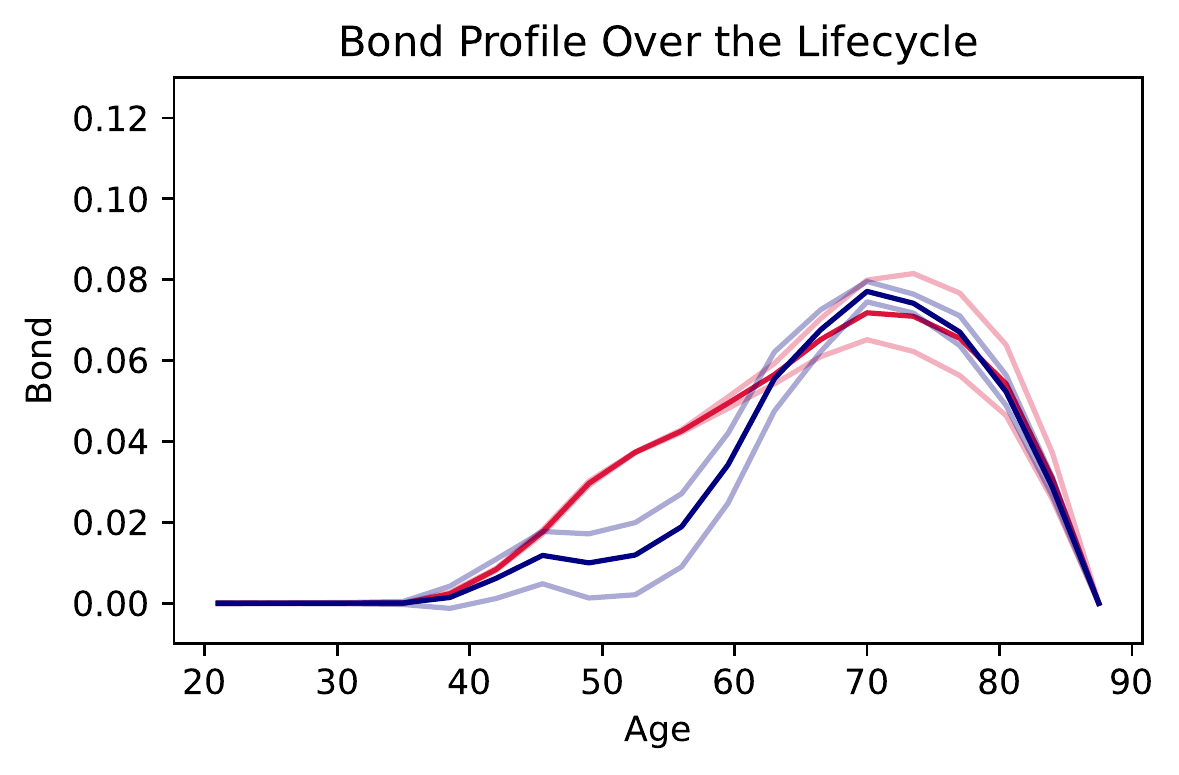}
\includegraphics[width = 0.32\textwidth]{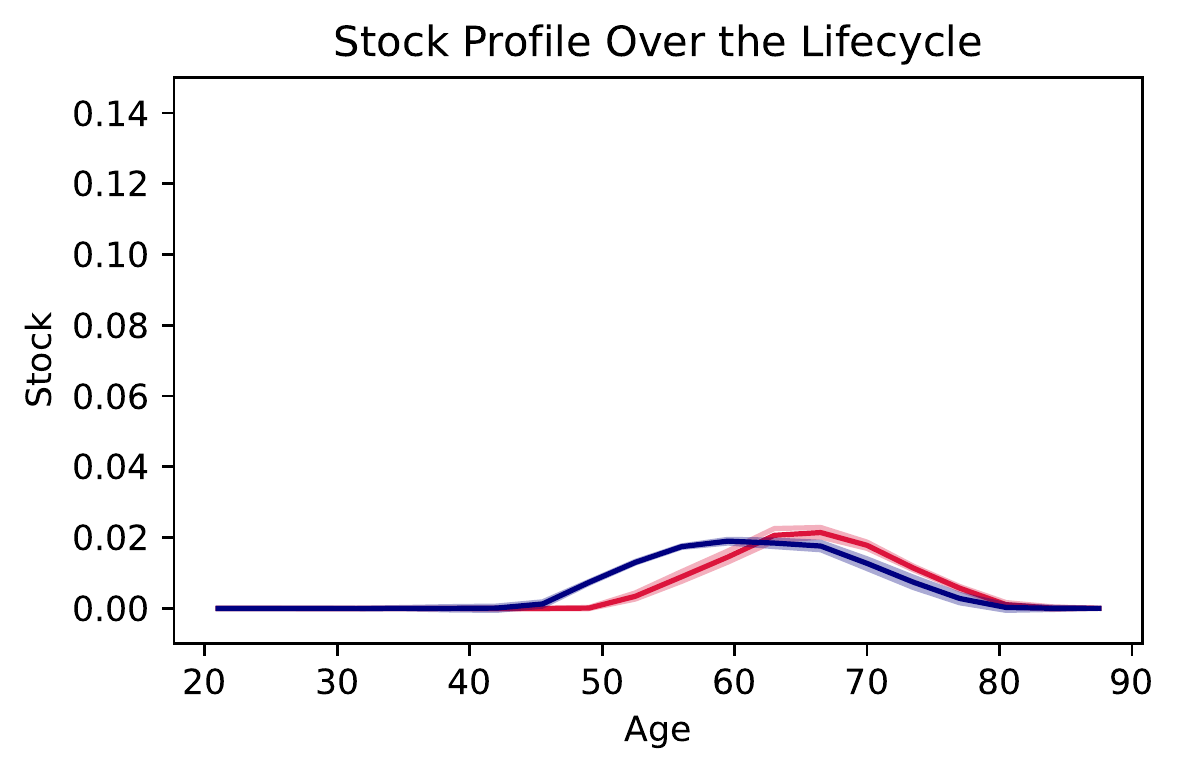}
\includegraphics[width = 0.32\textwidth]{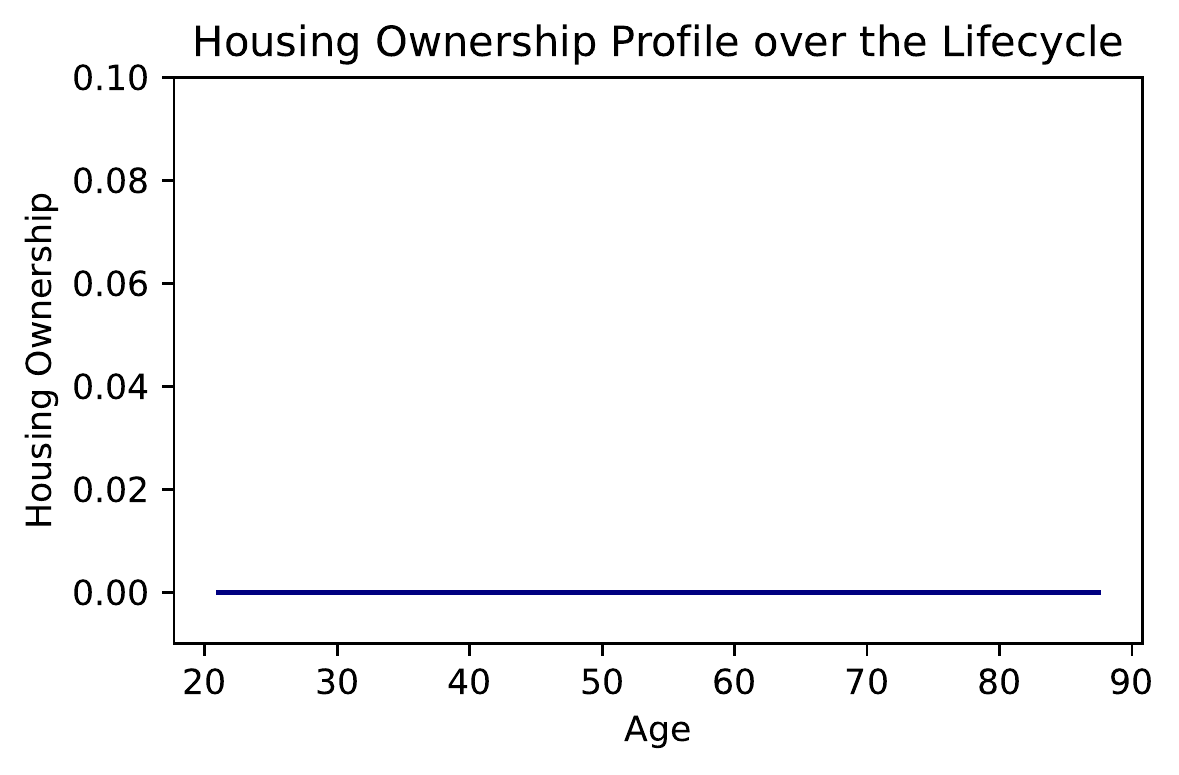}\\
\includegraphics[width = 0.32\textwidth]{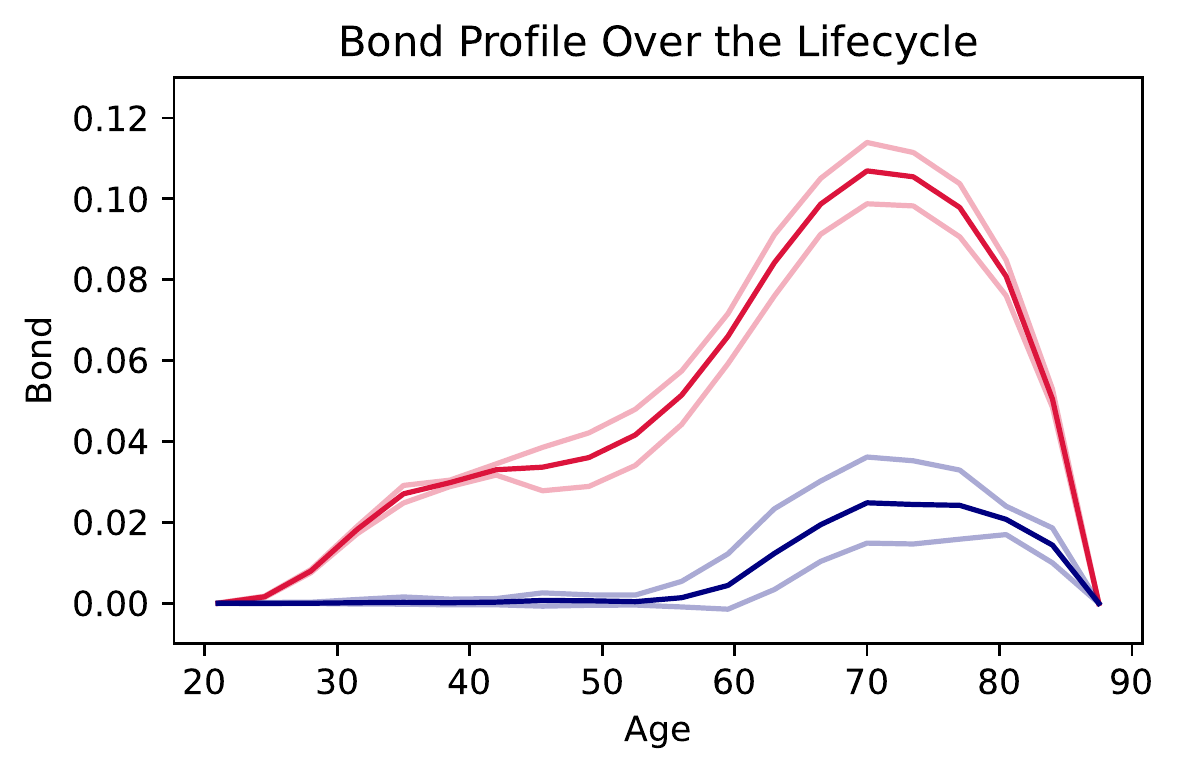}
\includegraphics[width = 0.32\textwidth]{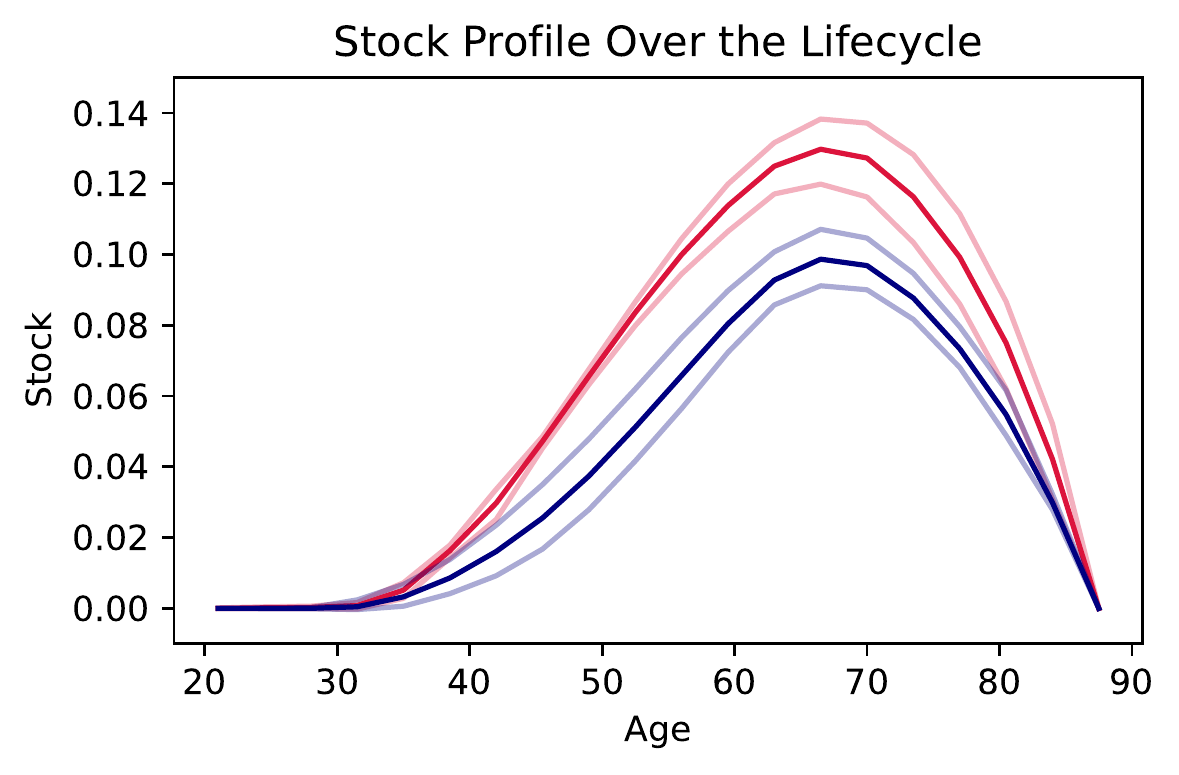}
\includegraphics[width = 0.32\textwidth]{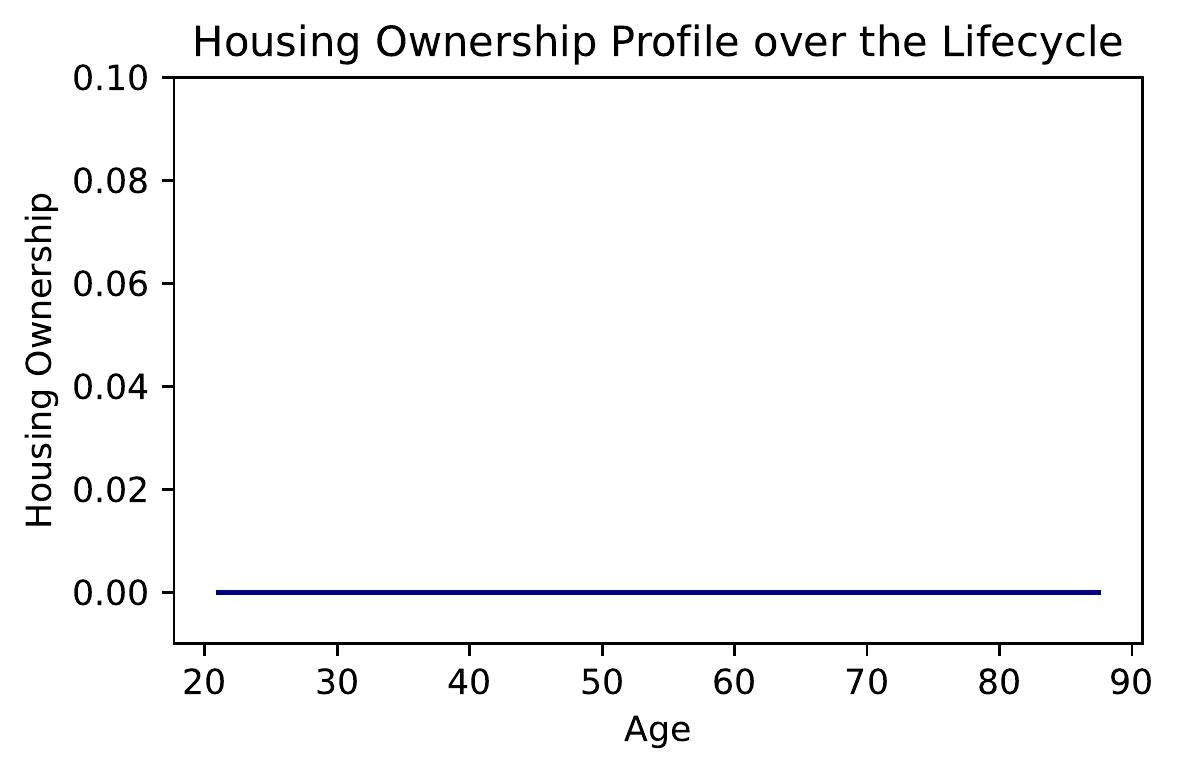}\\
\includegraphics[width = 0.32\textwidth]{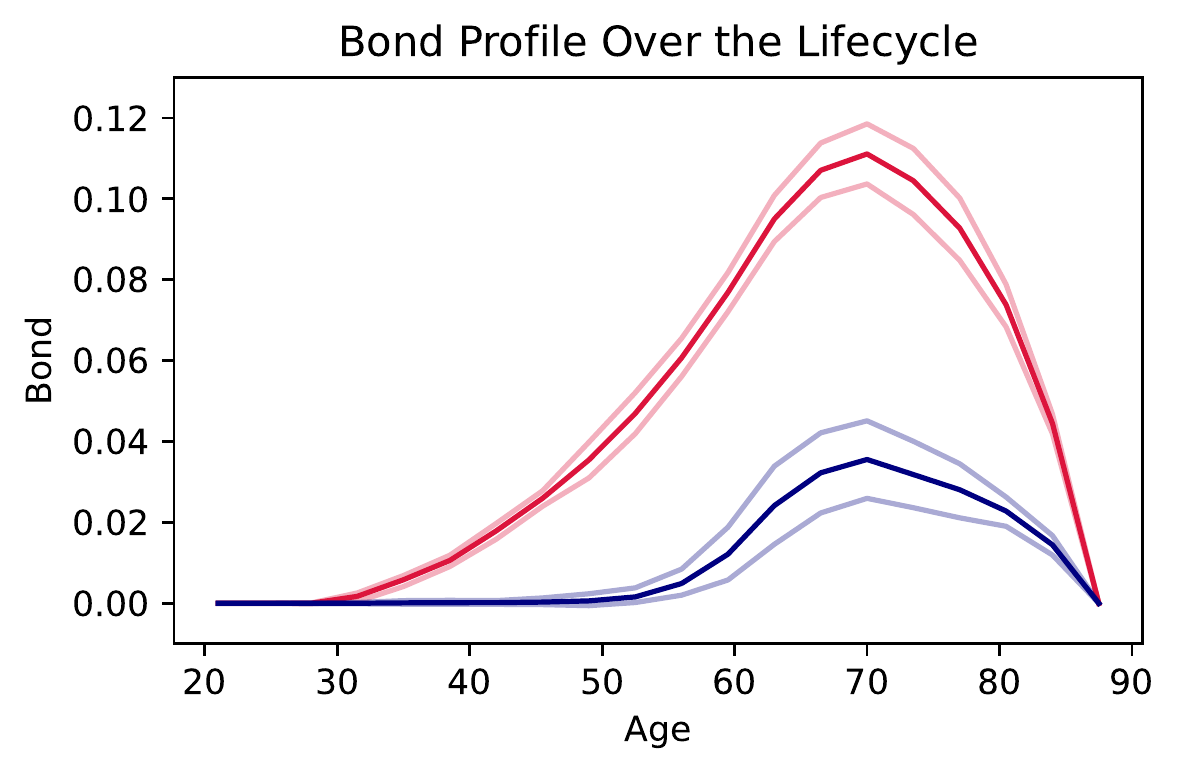}
\includegraphics[width = 0.32\textwidth]{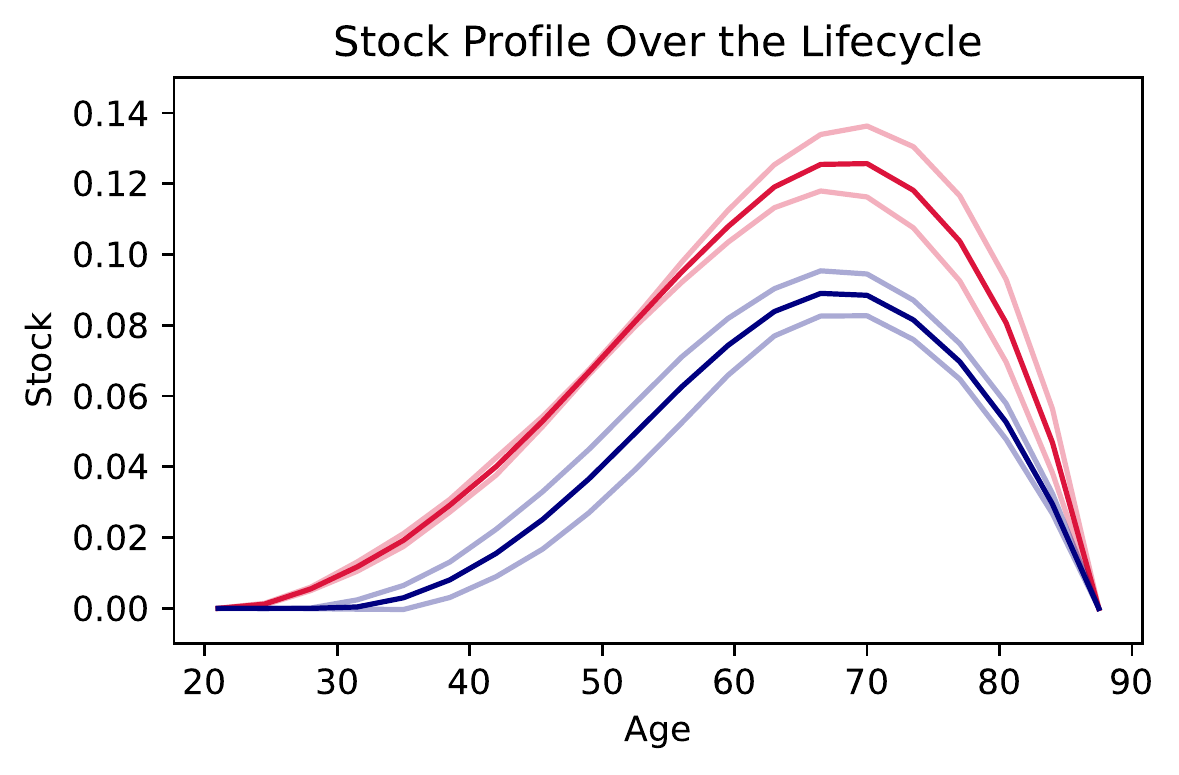}
\includegraphics[width = 0.32\textwidth]{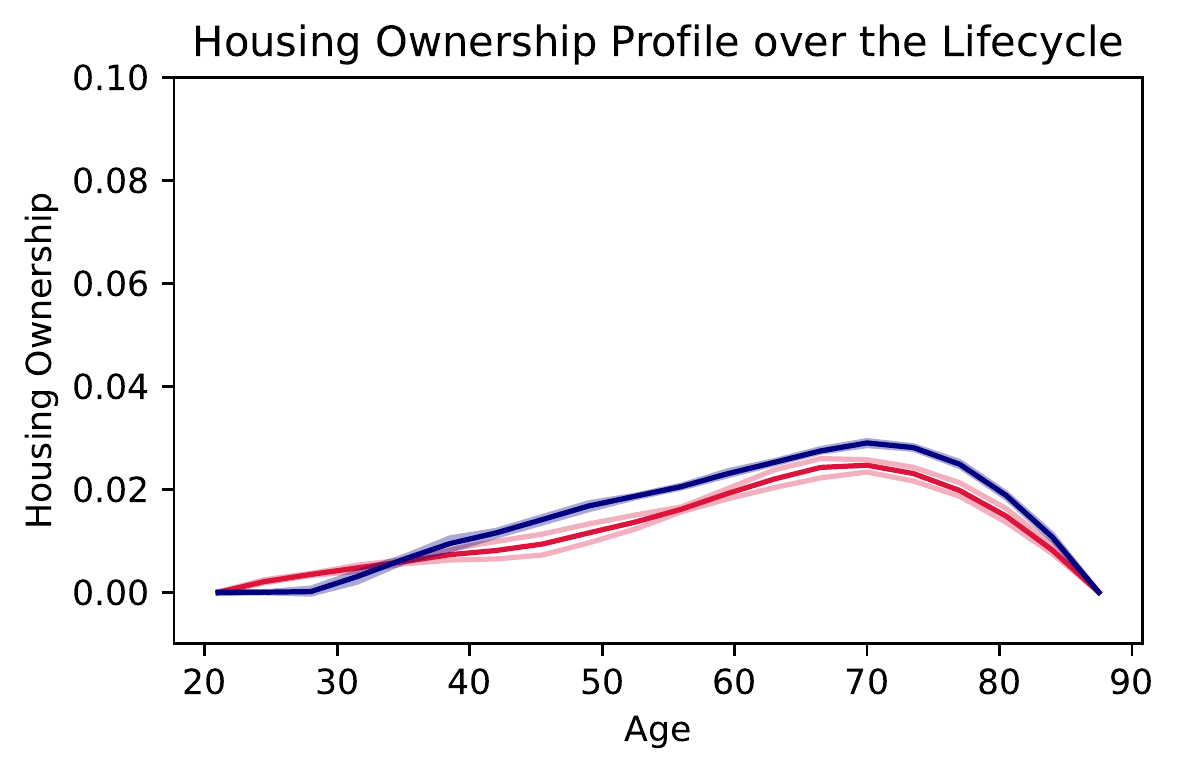}\\
\includegraphics[width = 0.32\textwidth]{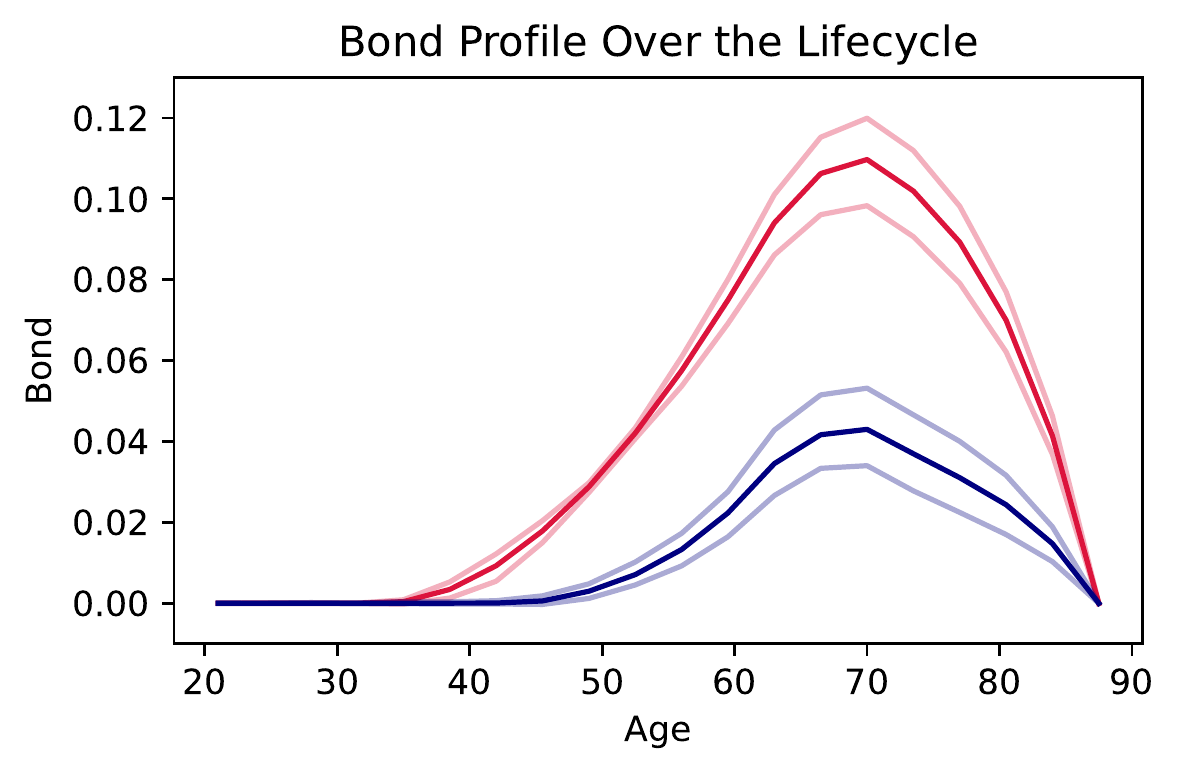}
\includegraphics[width = 0.32\textwidth]{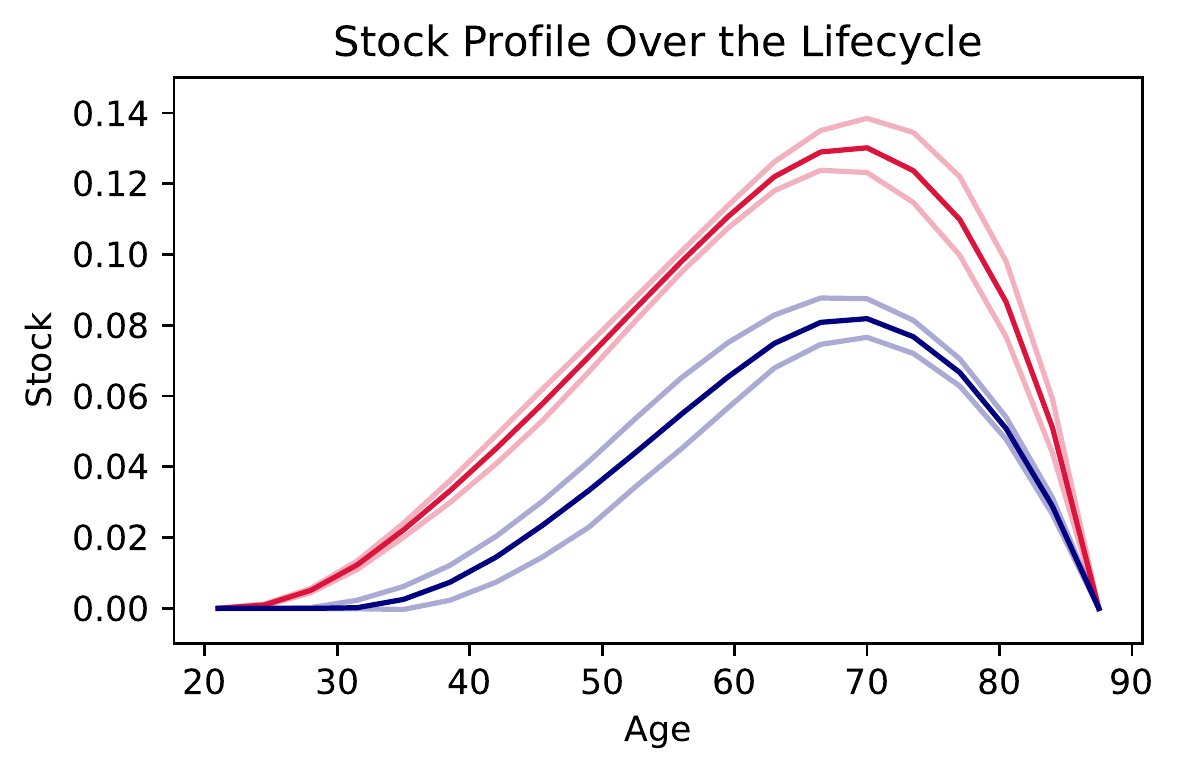}
\includegraphics[width = 0.32\textwidth]{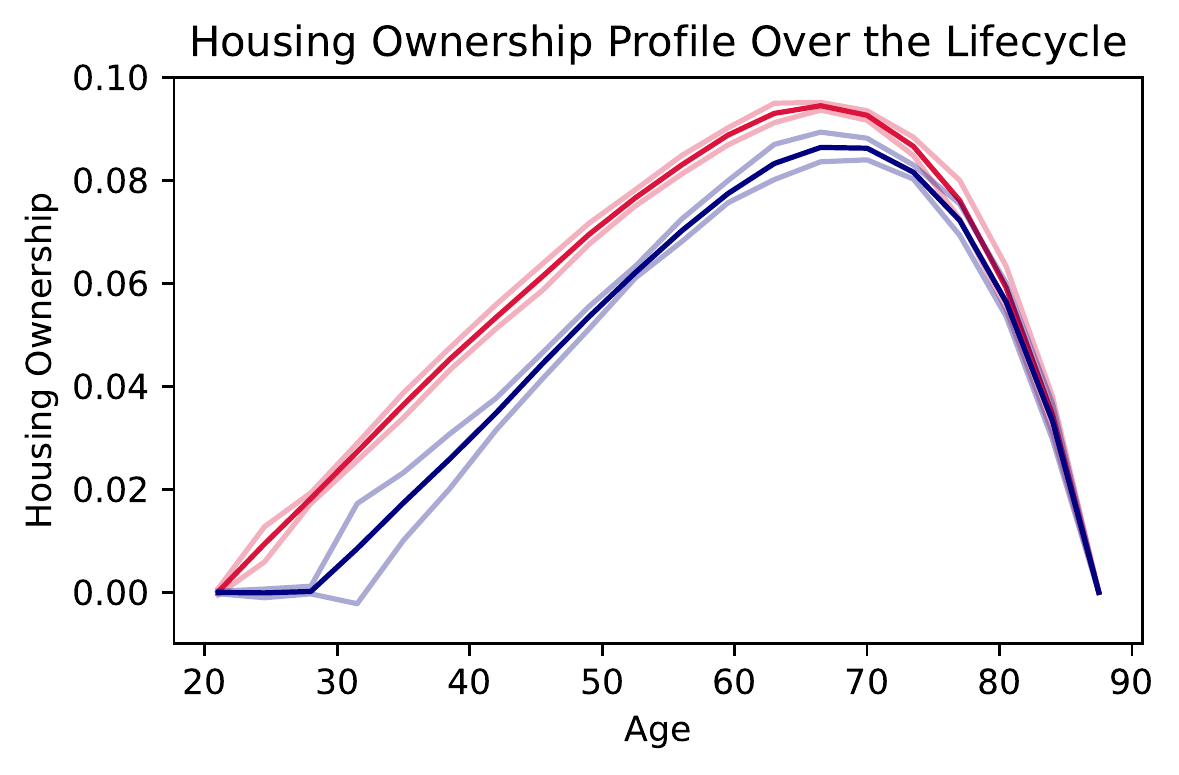}
\caption{Policy functions for the different assets at different stages of trainig of the homotopy algorithm.\label{fig:progress_homotopy}}
\end{figure}

\begin{figure}
\centering
\includegraphics[width = 0.32\textwidth]{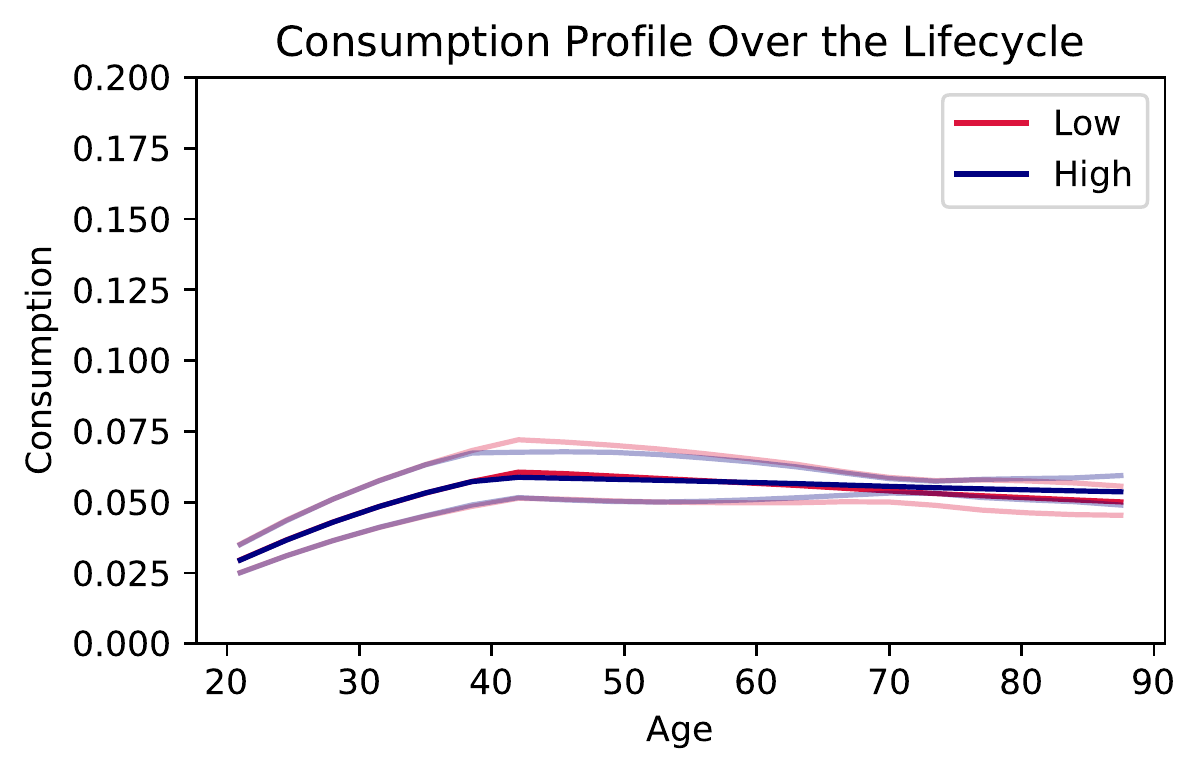}
\includegraphics[width = 0.32\textwidth]{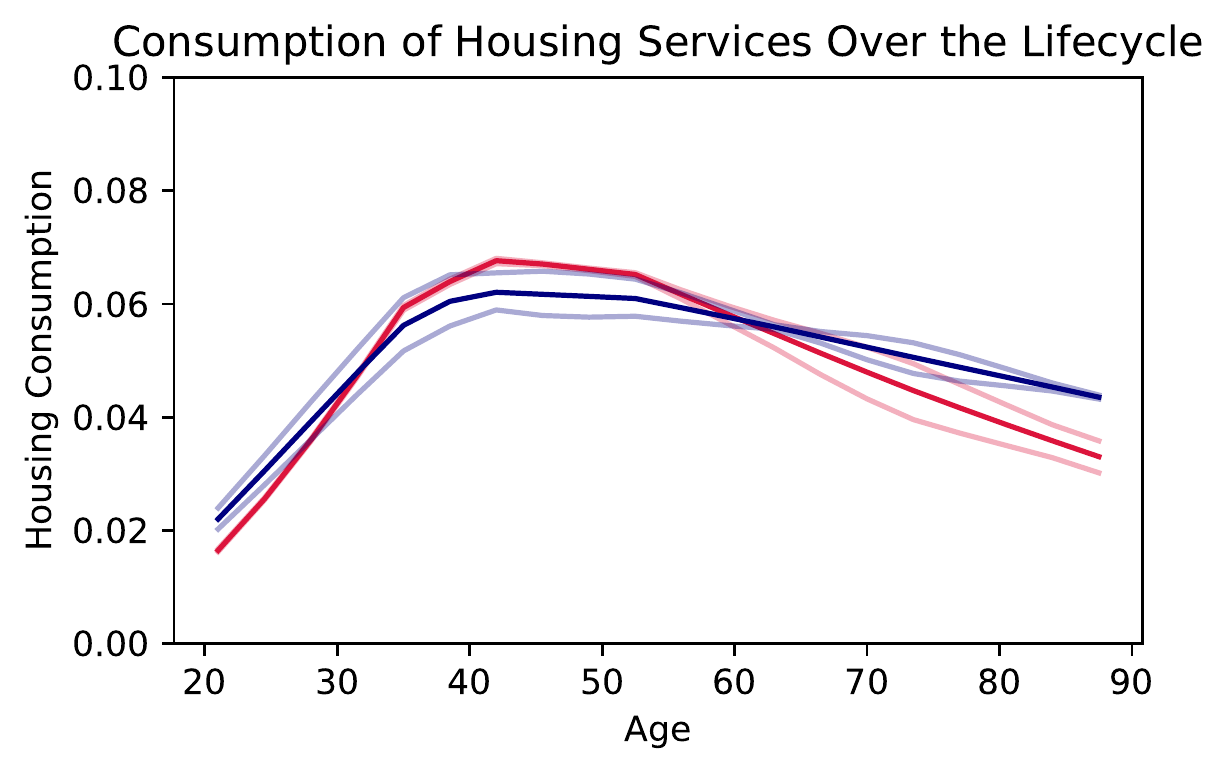}\\
\includegraphics[width = 0.32\textwidth]{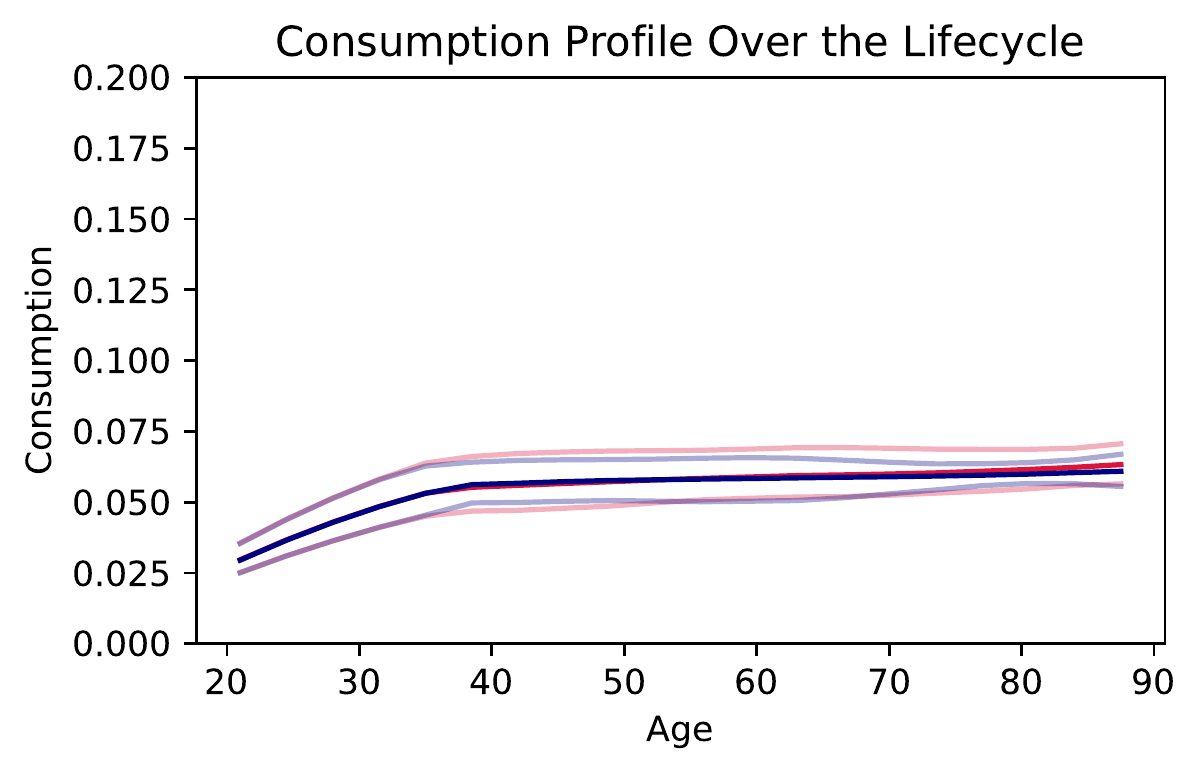}
\includegraphics[width = 0.32\textwidth]{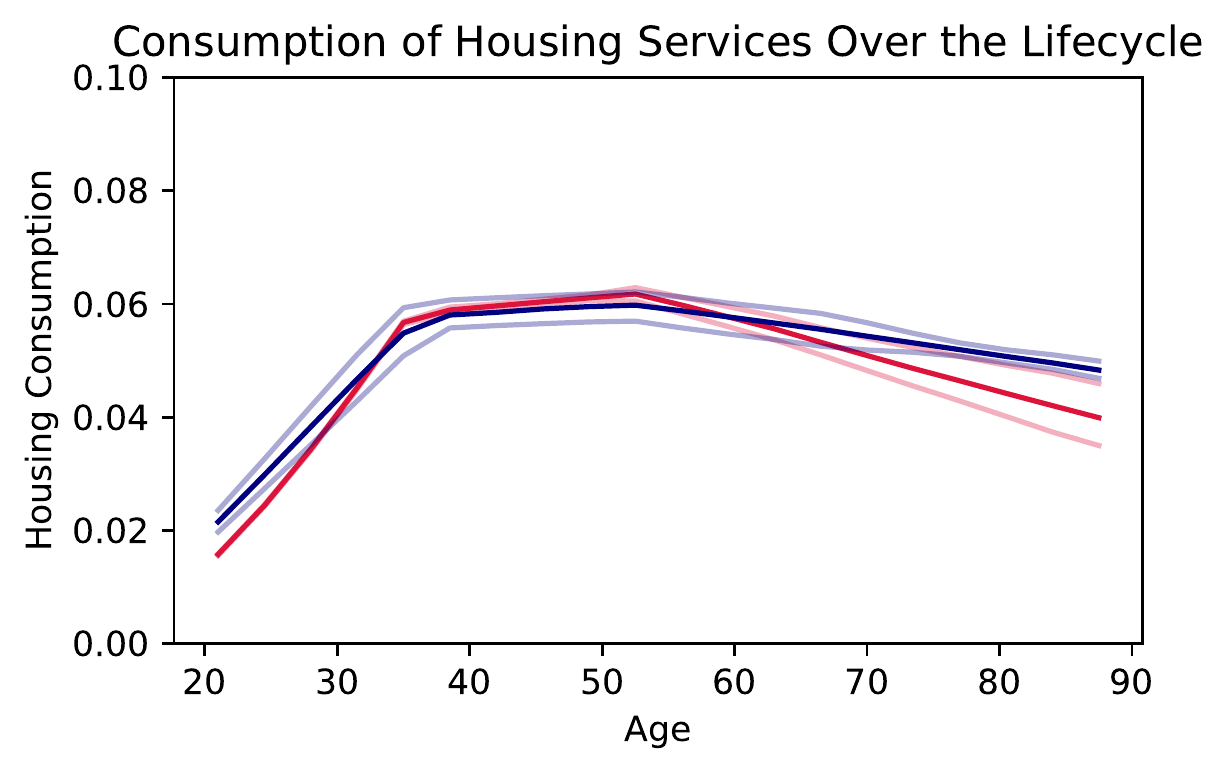}\\
\includegraphics[width = 0.32\textwidth]{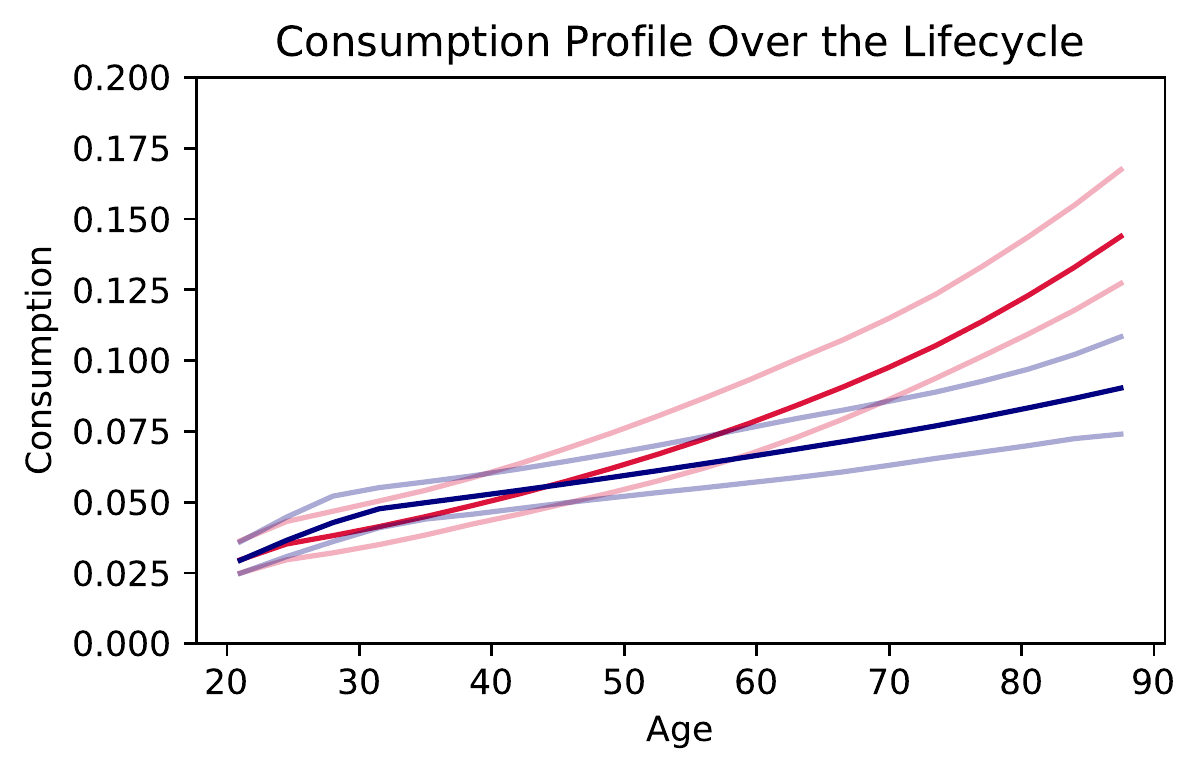}
\includegraphics[width = 0.32\textwidth]{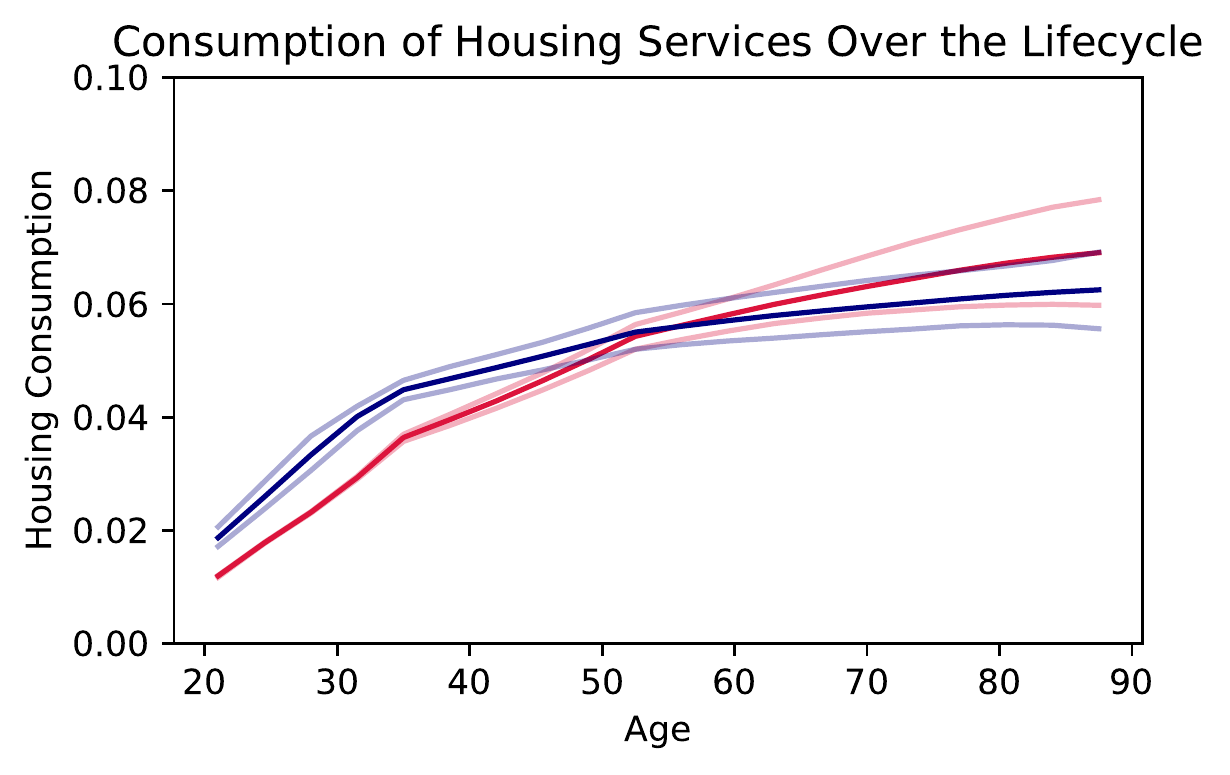}\\
\includegraphics[width = 0.32\textwidth]{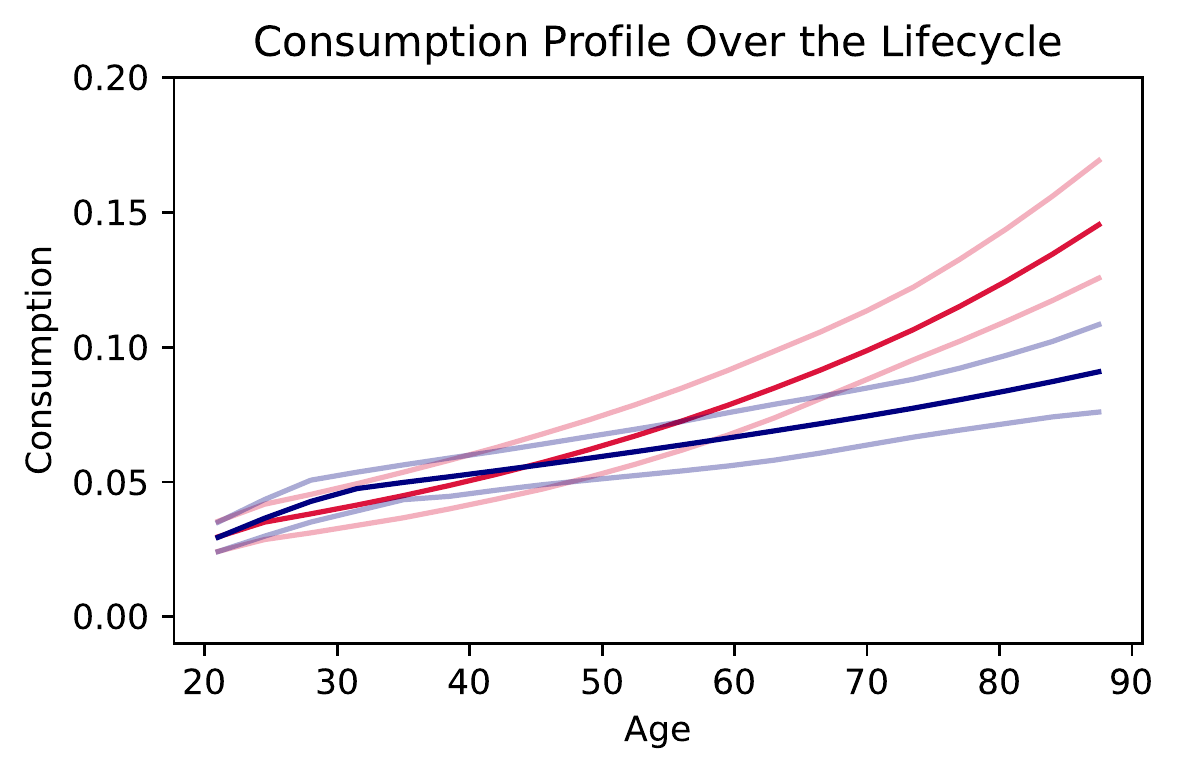}
\includegraphics[width = 0.32\textwidth]{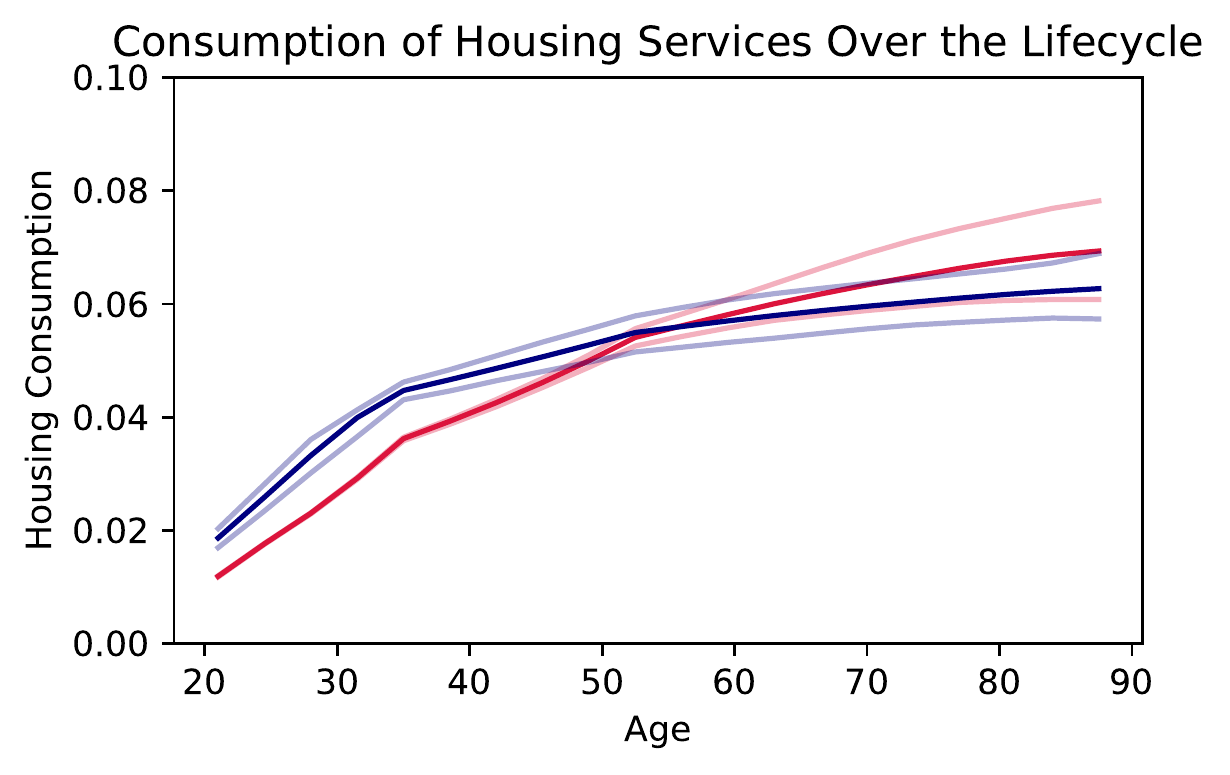}\\
\includegraphics[width = 0.32\textwidth]{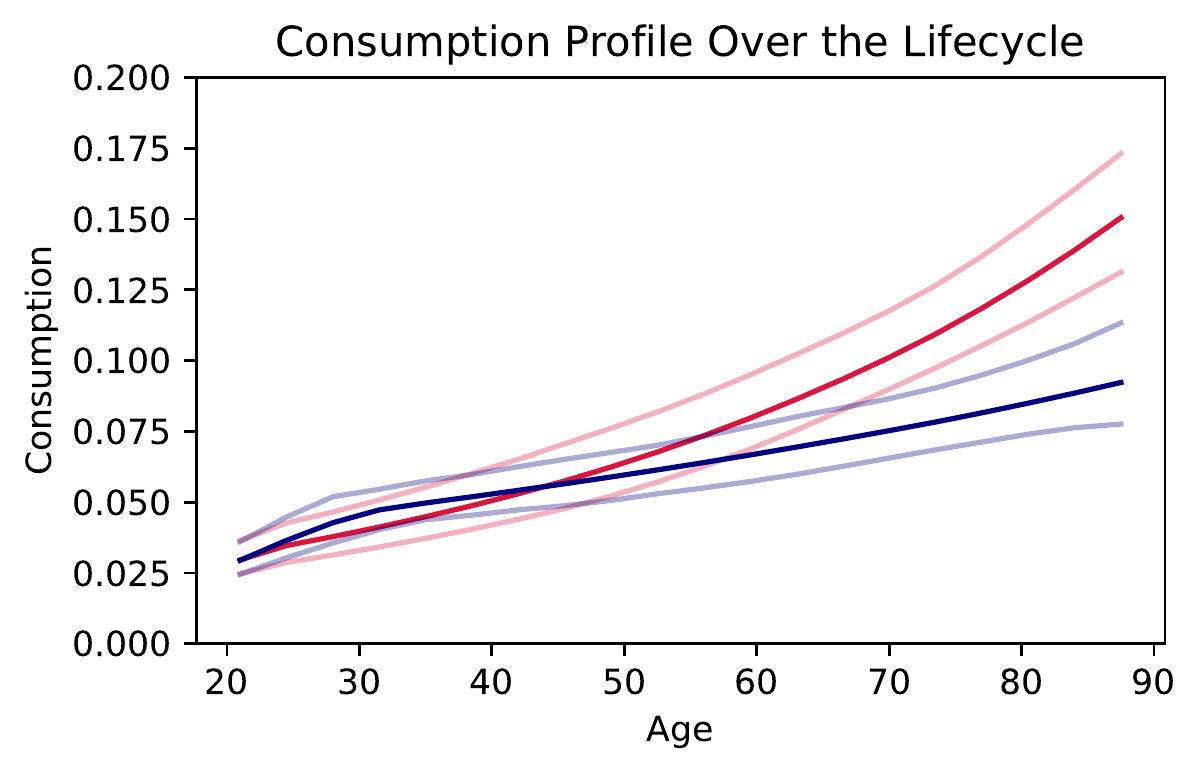}
\includegraphics[width = 0.32\textwidth]{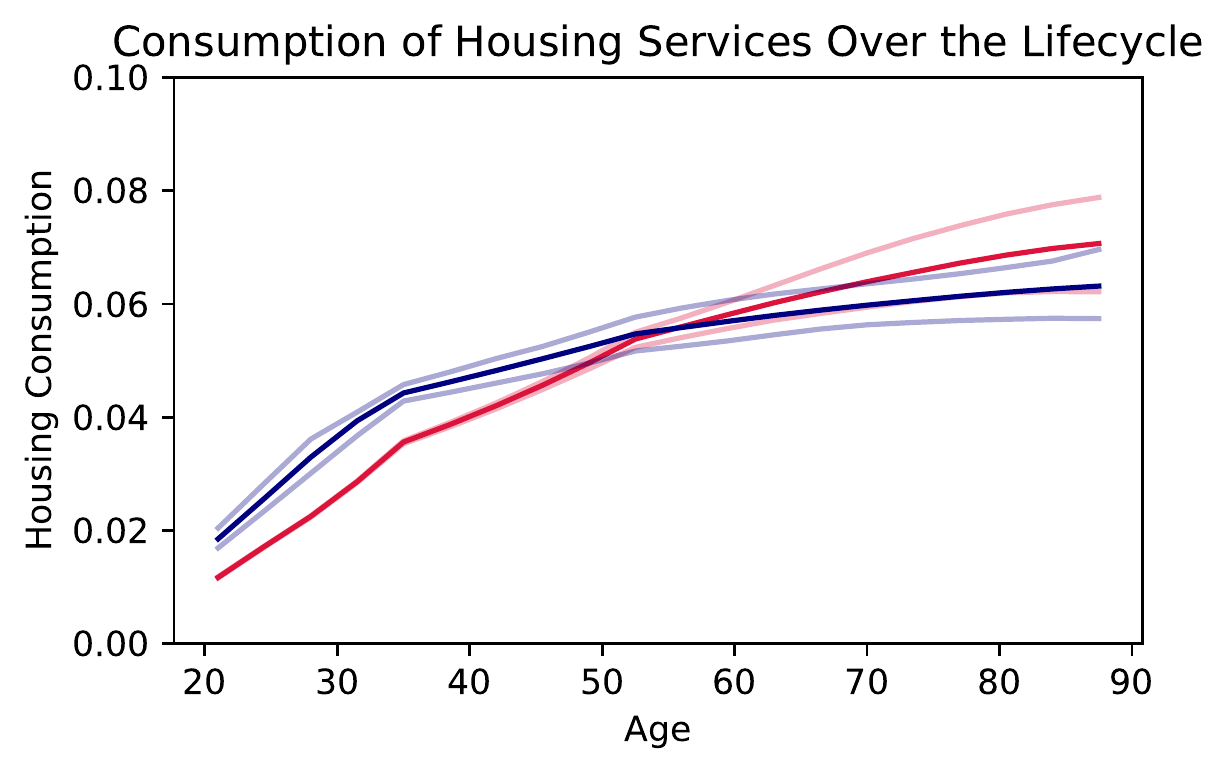}
\caption{Policy functions for consumption and the renting of housing services at different stages of training of the homotopy algorithm.\label{fig:progress_homotopy_cons}}
\end{figure}

Figure \ref{fig:progress_homotopy} shows the asset policies during different stages of our homotopy training algorithm.
Figure \ref{fig:progress_homotopy_cons} shows the corresponding policies for consumption and the renting of housing services.
The top panel in figure \ref{fig:progress_homotopy} shows the trained bond policy for each risk aversion type and age group, when the policies for stock and house ownership are still masked and the corresponding terms in the loss function are weighted with zero.
The second panel shows the policies, when a small amount of stock is introduced to the model, the policies for house ownership are still masked and equal to zero.
The third panel shows the policies, when a small amount of house ownership is introduced to the model.
Finally, the last panel shows the household policies, when the full amount of house ownership is introduced to the economy.
As the figures illustrate, the policies remain stable throughout the training.

Figure \ref{fig:progress_homotopy_cons} the corresponding profiles for consumption and the renting of housing services over the life-cycle during the stages of the homotopy algorithm. Increasing the supply of assets allows for a smoother consumption profile over the life-cycle.
In the top panel, which shows the economy with only a risk-free bond, the consumption profile over the life-cycle is slightly decreasing for unconstrained agents due to the low interest rate.
As more assets are added to the economy, the bond price drops and the interest rate rises.
In the bottom panel, the higher interest rate leads to an increasing consumption profile over the life-cycle.

Comparing the households with high risk aversion to the households with low risk aversion, figure \ref{fig:progress_homotopy_cons} shows that households with high risk aversion increase their consumption and save less early in life, at the cost of less consumption later in life.
For time separable expected utility with constant relative risk aversion, the risk aversion also pins down the intertemporal elasticity of substitution. 
The households with higher risk aversion are hence less willing to substitute consumption between periods and choose a more stable consumption sequence over their life-cycle.

\subsubsection{Accuracy}
\begin{figure}
\centering
\includegraphics[width=0.45\textwidth]{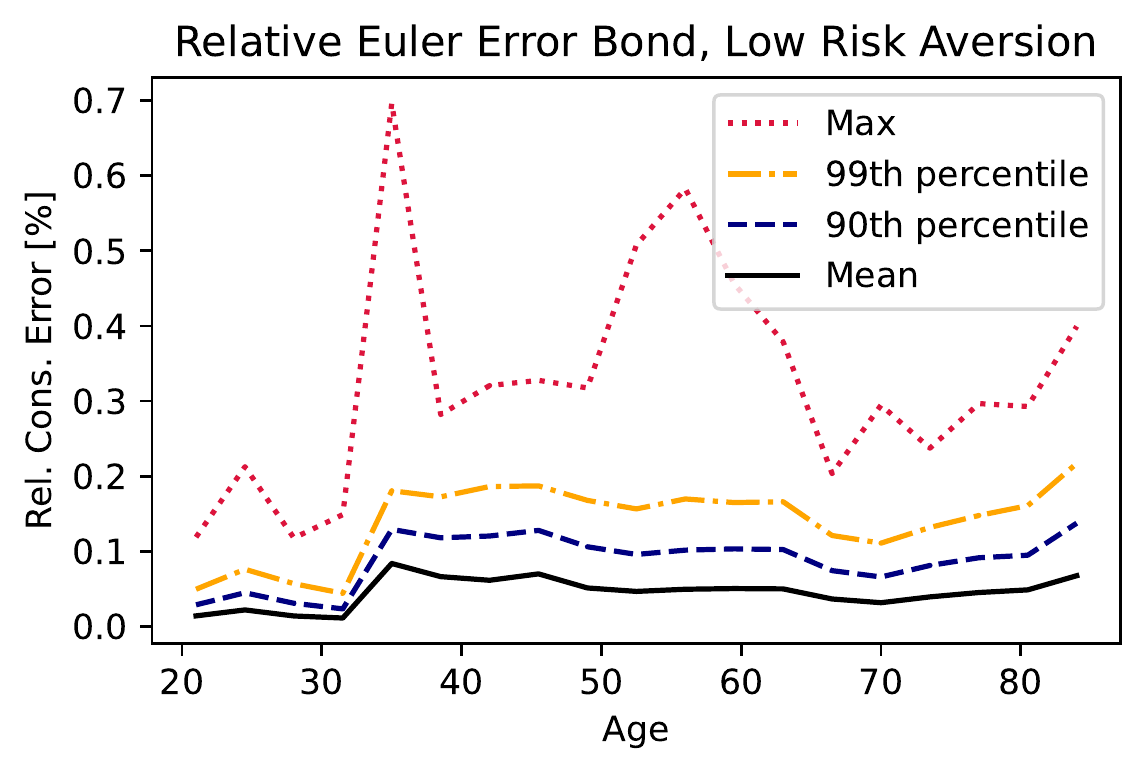}
\includegraphics[width=0.45\textwidth]{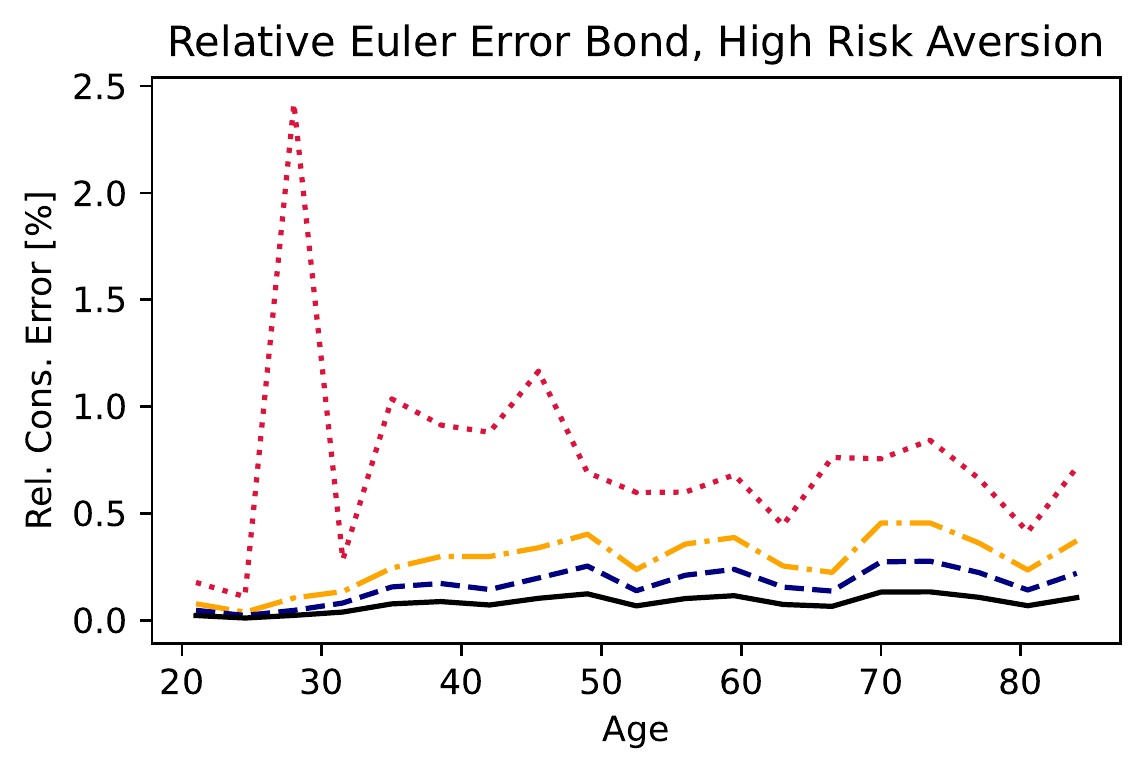}\\
\includegraphics[width=0.45\textwidth]{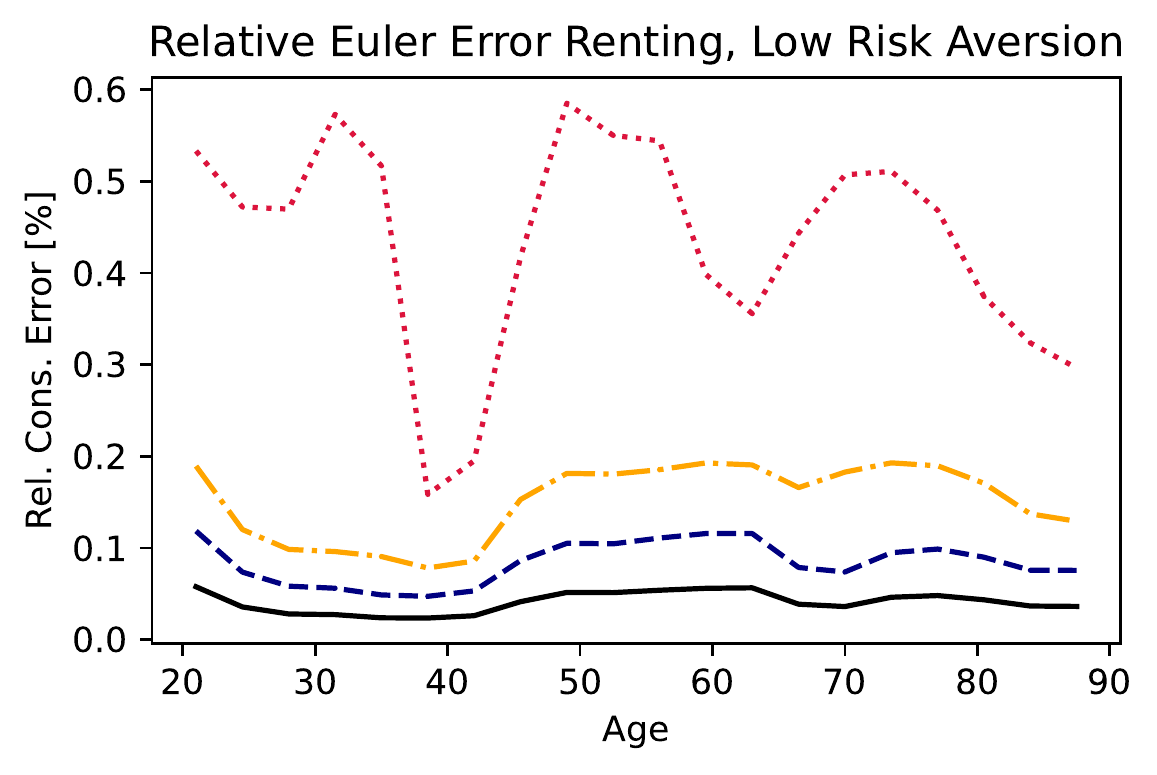}
\includegraphics[width=0.45\textwidth]{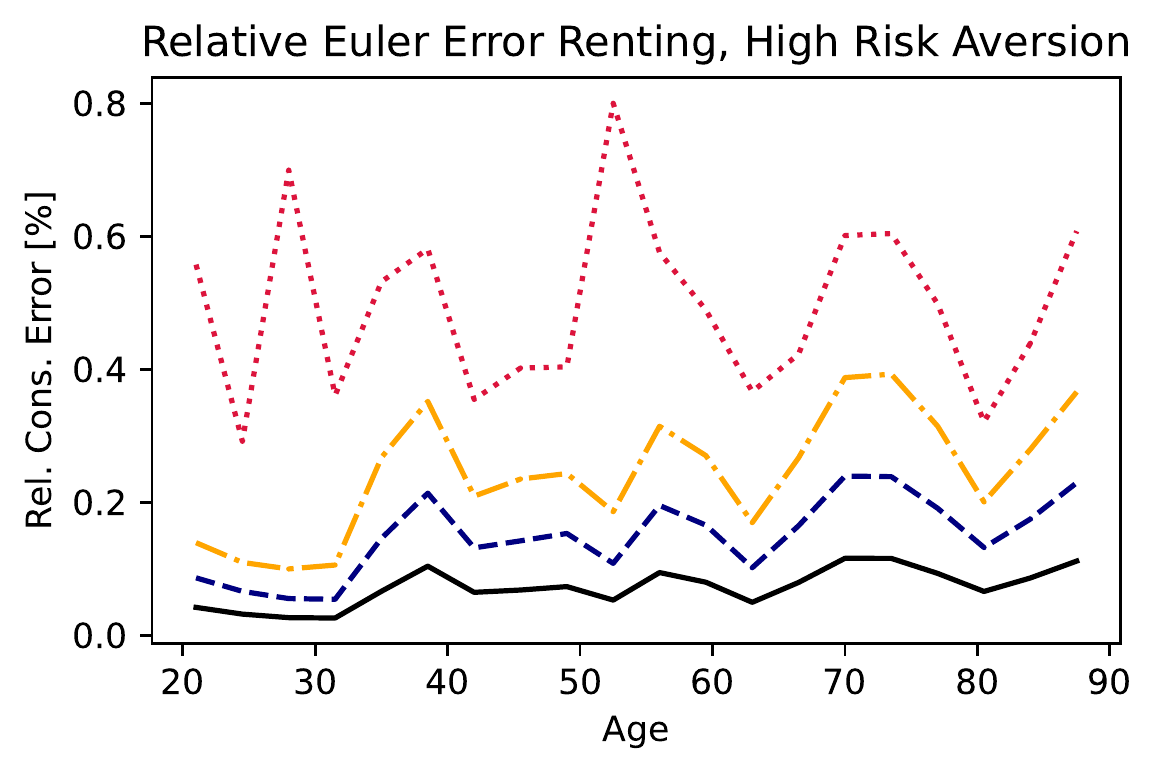}\\
\includegraphics[width=0.45\textwidth]{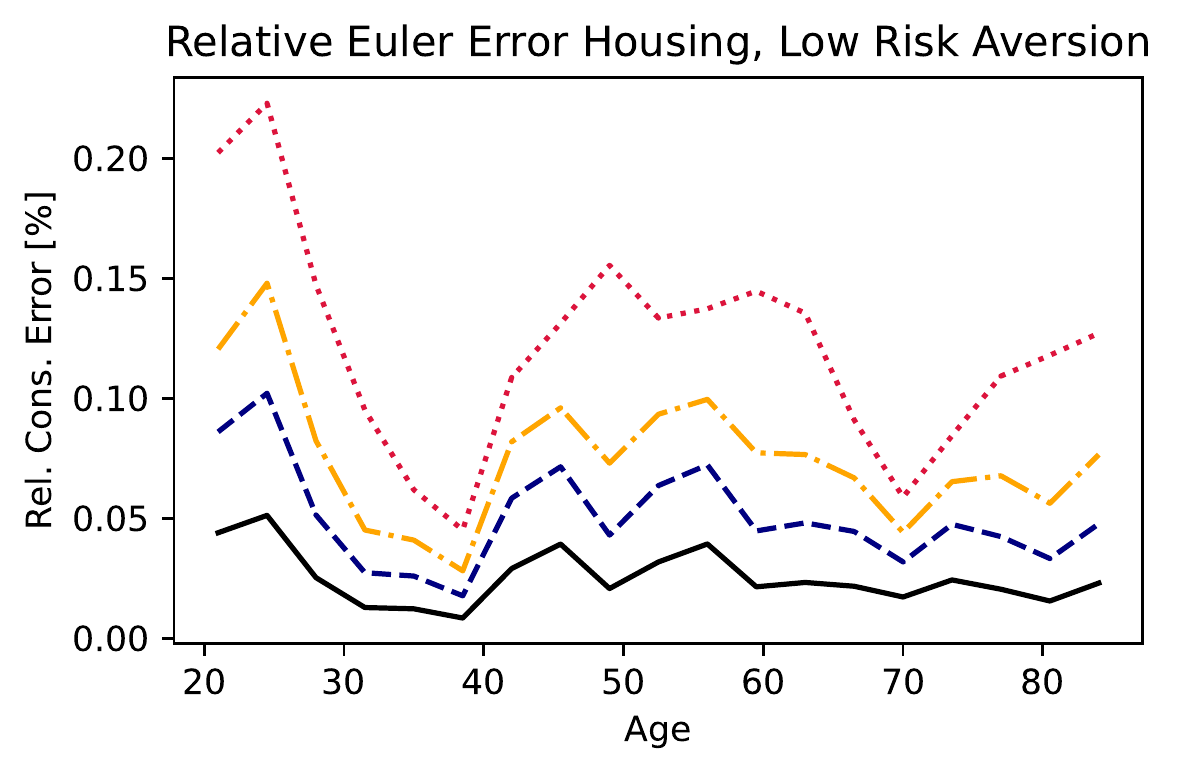}
\includegraphics[width=0.45\textwidth]{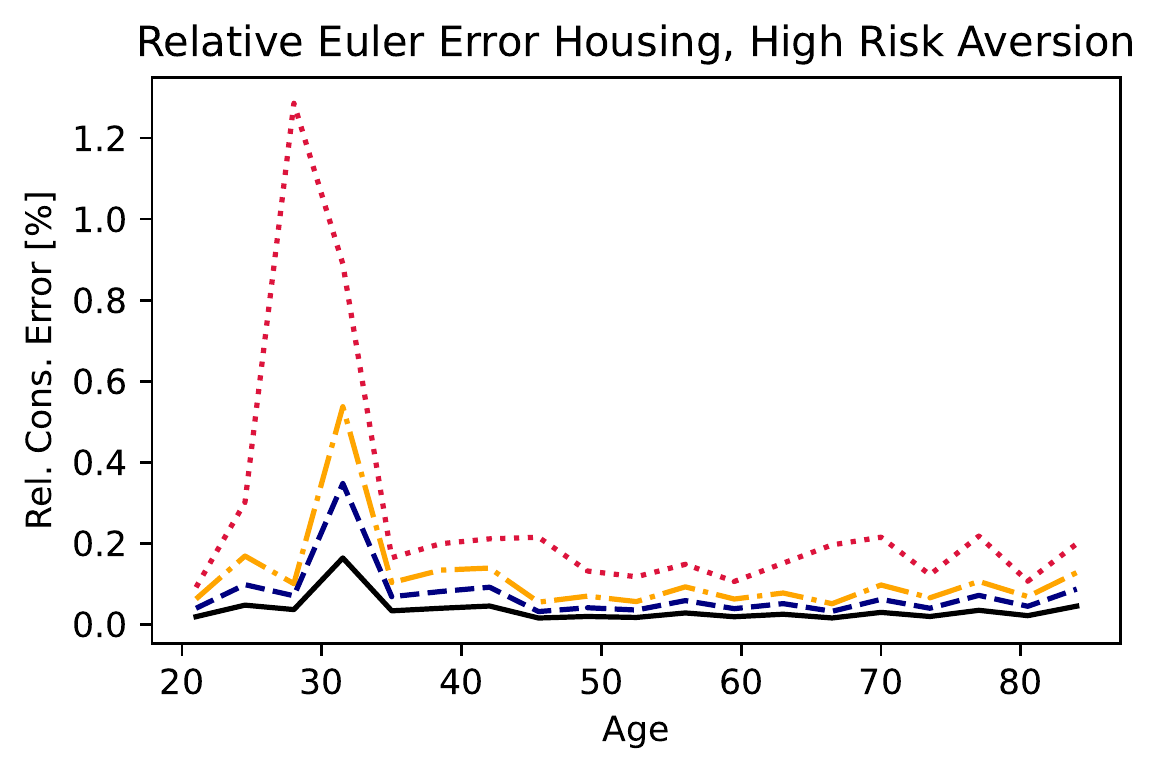}\\
\includegraphics[width=0.45\textwidth]{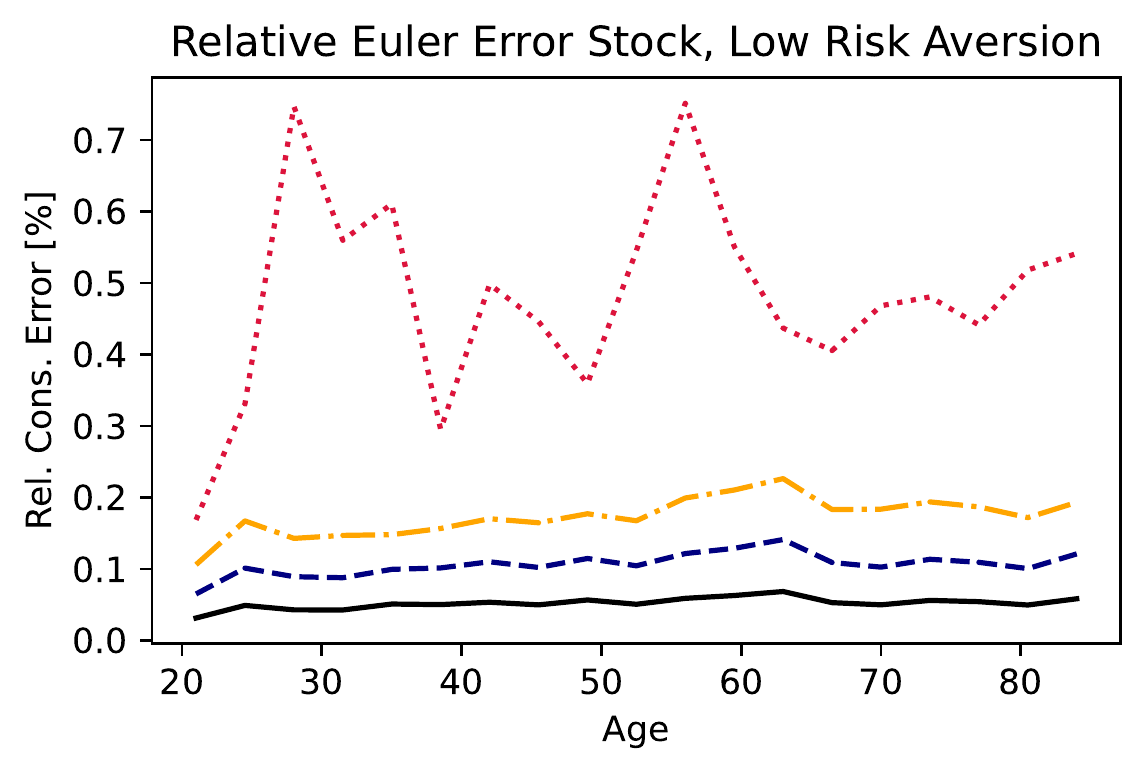}
\includegraphics[width=0.45\textwidth]{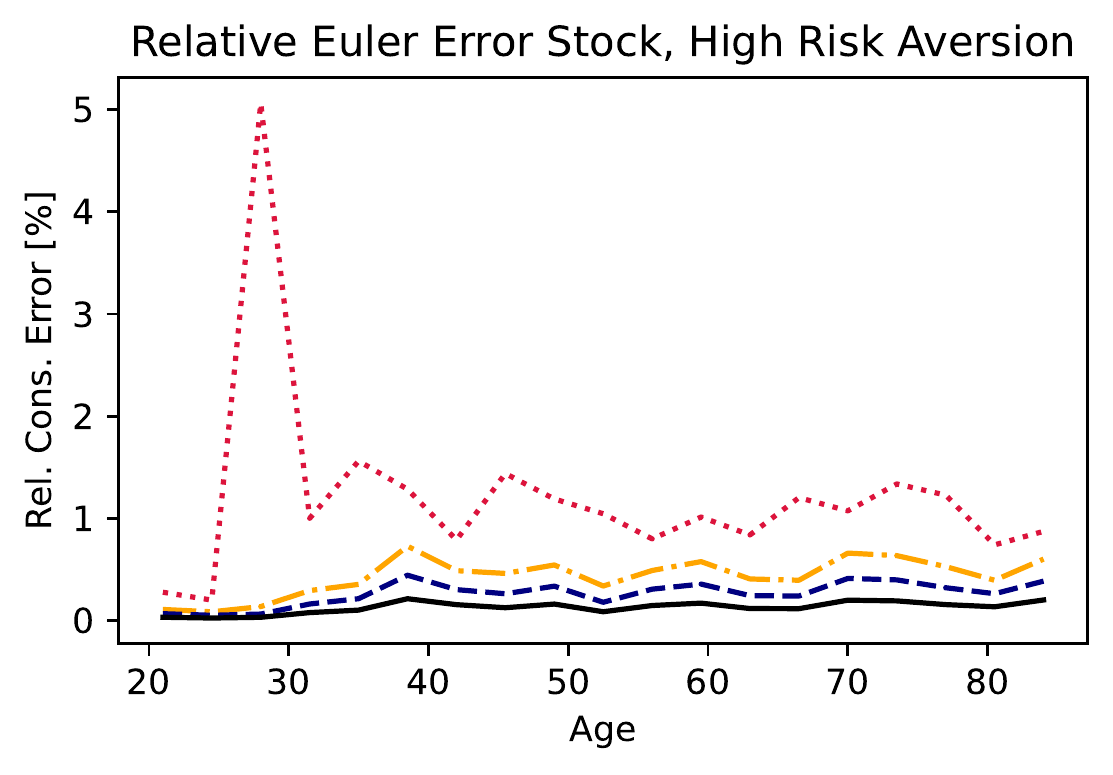}\\
\caption{\label{fig:het_final}Statistics on the errors in the remaining equilibrium conditions after training, expressed in \% and as relative consumption errors. 
The black solid line shows the mean error by age group, the blue dashed line shows the 90th percentile, the yellow dash-dotted line shows the 99th percentile and the dotted red line shows the maximum error by age group.
The left column shows the errors for the households with low risk aversion, the right panel shows the errors for the households with high risk aversion.
The top row shows the errors in the optimality conditions for bond purchses, the second row shows the errors in the equilibrium conditions for the choice of housing services, the third row shows the errors for the equilibrium conditions for house ownership, and the last row shows the errors in the equilibrium conditions for the stock choice.
}
\end{figure}

To assess the accuracy of our solution we compute the errors in the equilibrium conditions on an approximate ergodic set of the economy. To obtain the approximate ergodic set, we simulate the dataset of $N^{\text{trajectories}} = 8192$ states, generated by the last step of our training procedure, forward without retraining the neural network.\footnote{We simulate the final dataset $256$ periods forward without further training to test the ``out-of-sample'' performance of our approximate solution.} As shown in the figure \ref{fig:het_final}, the remaining errors are low, with the 99th percentile well below $1\%$ for each age group and optimality condition.
\section{Conclusion}
We make two distinct but complementary contibutions in the context of deep learning based solution methods. First, we show how market clearing and borrowing constraints can be encoded directly into the architecture of the neural network. 
Second, we present a homotopy algorithm for solving portfolio choice problems. 
The idea of our homotopy algorithm is to start solving a single asset and then add additional assets one by one and to slowly increase their supply in the economy.

\clearpage

\bibliography{Bibliography}{}
\bibliographystyle{apalike}

\clearpage

\renewcommand{\appendixpagename}{Appendix}
\begin{appendices}
\section{Parameters for the single asset model\label{sec:params_single}}
\begin{figure}
\centering
\includegraphics[width = 0.49\textwidth]{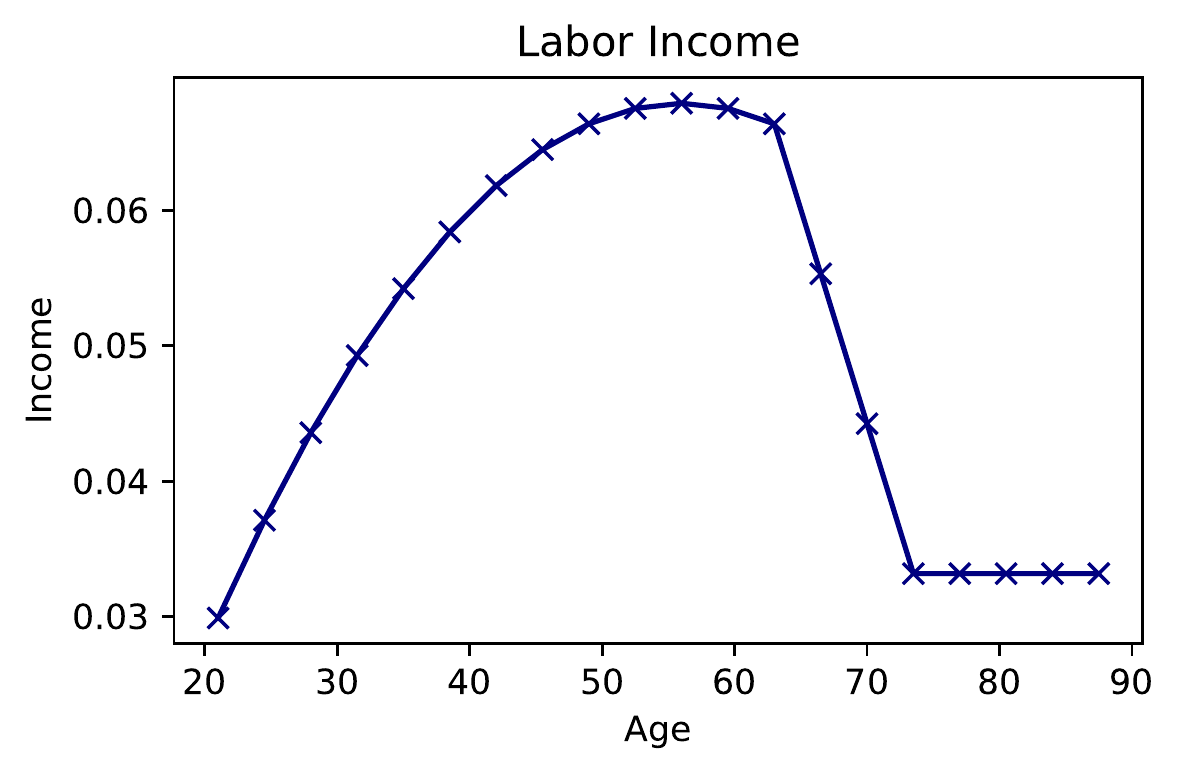}
\includegraphics[width = 0.49\textwidth]{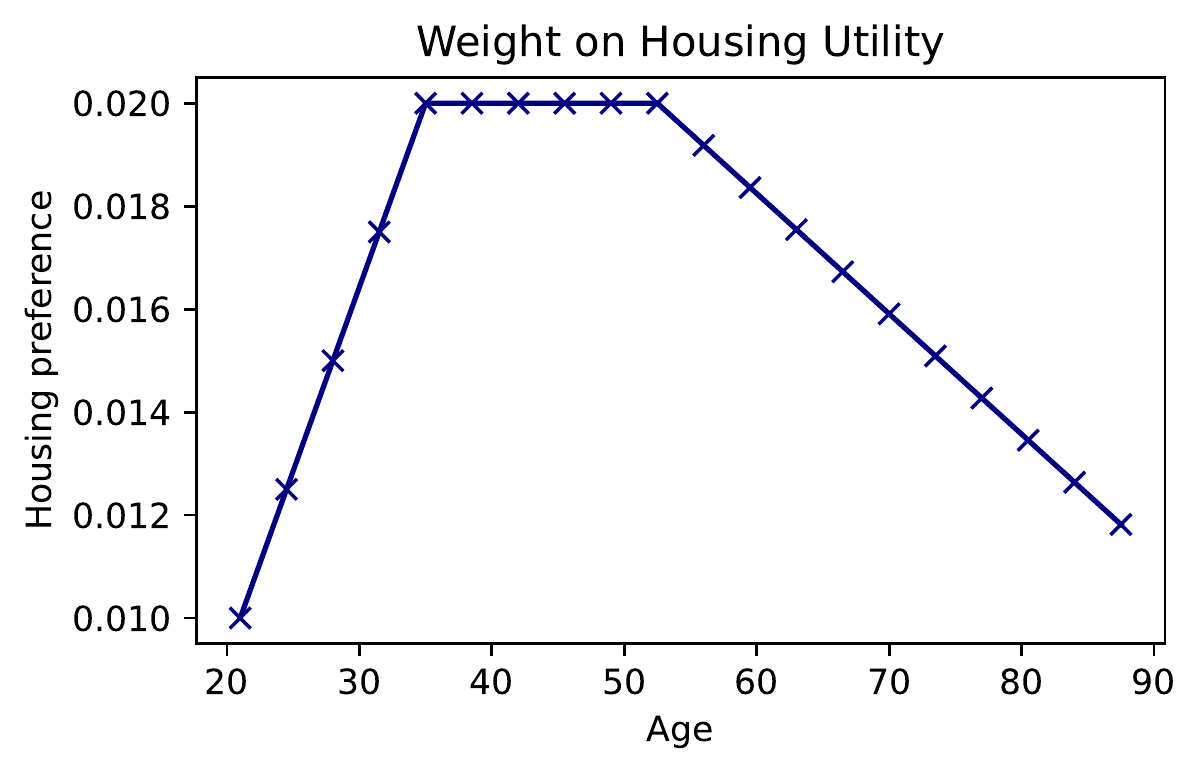}
\caption{\label{fig:inc_calib} The left panel shows the claim to aggregate labor income over the lifecycle. The right panel shows the age-specific weight on the utility from renting housing services.}
\end{figure}

We model $H=20$ age groups, thinking of one model period as roughly 3.5 years. We choose the patience $\beta$ to match a yearly patience of 0.96 and the $\rho$ and $\sigma$ to match a yearly persistence of $0.8$ and standard deviation of innovation $0.03$. We choose an adjustment cost parameter $\zeta^b = 0.5$ and a coefficient of relative risk aversion $\gamma = 3$.
The borrowing constraint is set to $\underline{b} = 0$.
The parameter values are summarized in table \ref{tab:calib_simple}.
\begin{table}[tb!]
\begin{center}
\begin{tabular}{ccccccccccc}
\toprule 
Parameters & $H$ & $y^h$ & $\psi^h$ & \underline{h} & $\beta$ & $\gamma$ & $\underline{b}$ & $\zeta^b$ & $\rho$ & $\sigma$ \\
\midrule
Values & 20 & see figure \ref{fig:inc_calib} & see figure \ref{fig:inc_calib} & $5\times10^{-5}$ & 0.867 & 3 & 0 & 0.5 & 0.458 & 0.043\\
\bottomrule
\end{tabular}
\caption{Parameter values chosen for the single asset model.\label{tab:calib_simple}}
\end{center}
\end{table}
The life-cycle profile of labor income is chosen as in \cite{azinovic2022deep} and shown in the left panel in figure \ref{fig:inc_calib}.
Similarly, the life-cycle profile of the preference for housing services is shown in the right panel in figure \ref{fig:inc_calib}.

\section{Household problem in the economy with multiple assets\label{app:hh_prob}}
In the following we detail the household model described in section \ref{sec:multiasset_model} and provide the KKT conditions for optimality.
As before, each set of KKT conditions can encoded into a single Fischer-Burmeister function.
\begin{align}
V^h(z_t, b_t^h, s_t^h, h^{o, h}_{t}) &= \max_{h^h_{r,t},b_{t+1}^{h+1}, s_{t+1}^{h+1}, h^{o, h+1}_{t+1}}\left\{u(c_t^h) + \psi^h v(h^{r, h}_{t}) + \beta \E{V^{h+1}(z_{t+1}, b_{t+1}^{h+1}, s_{t+1}^{h+1}, h^{o, h+1}_{t+1})}\right\}\\
c_t^h &= z_t y^h + b_t^h + s_t^h (p_t^s + d_t^s) + h^{o, h}_{t} (p_t^{o} + p_t^r) \nonumber \\
&- b_{t+1}^{h+1} p_t^b - s_{t+1}^{h+1} p_{t+1}^s - h^{o, h+1}_{t+1} p_t^{o} - h^{r, h}_{t}p_t^{r} \nonumber \\
&-p_t^b\zeta^b \frac{1}{2} \biground{b_{t+1}^{h+1} - b_{t}^{h}}^2-p_t^s\zeta^s \frac{1}{2} \biground{s_{t+1}^{h+1} - s_{t}^{h}}^2-p_t^o\zeta^h \frac{1}{2} \biground{h_{t+1}^{o, h+1} - h_{t}^{o, h}}^2\\
\text{subject to}:& \nonumber \\
0 &\leq b_{t+1}^{h+1} - \underline{b}  \\
0 &\leq s_{t+1}^{h+1} - \underline{s}  \\
0 &\leq h_{t+1}^{o, h+1} - \underline{h}^o 
\end{align}
Karush Kuhn Tucker conditions for the bond are given by
\begin{align}
[b_{t+1}^{h+1}]&~~u'(c_t^h)\biground{p_t^b + p_t^b\zeta^b\biground{b_{t+1}^{h+1} - b_t^h}}= \beta \E{u'(c_{t+1}^{h+1})\biground{1 + p_{t+1}^b\zeta^b \biground{b_{t+2}^{h+2} - b_{t+1}^{h+1}}}} + \lambda_t^{b, h}  \\
\Leftrightarrow 0 &= \frac{\biground{u'}^{-1}\biground{\frac{\beta \E{u'(c_{t+1}^{h+1})\biground{1 + p_{t+1}^b\zeta^b \biground{b_{t+2}^{h+2} - b_{t+1}^{h+1}}}} + \lambda_t^{b, h}}{p_t^b + p_{t}^b\zeta^b\biground{b_{t+1}^{h+1} - b_t^h}}}}{c_t^h} - 1\\
[b_{t+1}^{h+1}]&~~0 \leq (b_{t+1}^{h+1} - \underline{b}) \\
[b_{t+1}^{h+1}]&~~0 \leq \lambda^{b, h}_t \\
[b_{t+1}^{h+1}]&~~0 = \lambda^{b, h}_t (b_{t+1}^{h+1} - \underline{b}).
\end{align}
Karush Kuhn Tucker conditions for the stock are given by
\begin{align}
[s_{t+1}^{h+1}]&~~u'(c_t^h)\biground{p_t^s + p_t^s\zeta^s\biground{s_{t+1}^{h+1} - s_t^h}}= \beta \E{u'(c_{t+1}^{h+1})\biground{p_{t+1}^{s+1} + d_{t+1} + p_{t+1}^s\zeta^s \biground{s_{t+2}^{h+2} - s_{t+1}^{h+1}}}} + \lambda_t^{s, h}\\
\Leftrightarrow 0 &= \frac{\biground{u'}^{-1}\biground{\frac{\beta \E{u'(c_{t+1}^{h+1})\biground{p_{t+1}^{s+1} + d_{t+1} + p_{t+1}^s\zeta^s \biground{s_{t+2}^{h+2} - s_{t+1}^{h+1}}}} + \lambda_t^{s, h}}{p_t^s + p_{t}^s\zeta^s\biground{s_{t+1}^{h+1} - s_t^h}}}}{c_t^h} - 1\\
[s_{t+1}^{h+1}]&~~0 \leq (s_{t+1}^{h+1} - \underline{s}) \\
[s_{t+1}^{h+1}]&~~0 \leq \lambda^{s, h}_t \\
[s_{t+1}^{h+1}]&~~0 = \lambda^{s, h}_t (s_{t+1}^{h+1} - \underline{s}).
\end{align}
Karush Kuhn Tucker conditions for owning the housing fund are given by
\begin{align}
[h_{t+1}^{o, h+1}]&~~u'(c_t^h)\biground{p_t^{o} + p_t^o\zeta^h\biground{h_{t+1}^{o, h+1} - h_t^{o,h}}}= \beta \E{u'(c_{t+1}^{h+1})\biground{p_{t+1}^{o} + p_{t+1}^{r} + p_{t+1}^o\zeta^h \biground{h_{t+2}^{o, h+2} - h_{t+1}^{o, h+1}}}} + \lambda_t^{o, h}\\
\Leftrightarrow 0&= \frac{\biground{u'}^{-1}\biground{\frac{\beta \E{u'(c_{t+1}^{h+1})\biground{p_{t+1}^{o} + p_{t+1}^{r} + p_{t+1}^o\zeta^h \biground{h_{t+2}^{o, h+2} - h_{t+1}^{o, h+1}}}} + \lambda_t^{o, h}}{p_t^{o} + p_t^o\zeta^h\biground{h_{t+1}^{o, h+1} - h_t^{o,h}}}}}{c_t^h} - 1\\
[h_{t+1}^{o, h+1}]&~~0 \leq (h_{t+1}^{o, h+1} - \underline{h}^o) \\
[h_{t+1}^{o, h+1}]&~~0 \leq \lambda^{o, h}_t \\
[h_{t+1}^{o, h+1}]&~~0 = \lambda^{o, h}_t (h_{t+1}^{o, h+1} - \underline{h}^o).
\end{align}
First order condition for renting
\begin{align}
[h_{t}^{r, h}]&~~u'(c_t^h)p_t^{r} = \psi^h v'(h_{t}^{r, h})\\
\Leftrightarrow 0 &= \frac{\biground{u'}^{-1}\biground{\frac{\psi^h v'(h_{t}^{r, h})}{p_t^{r}}}}{c_t^h} - 1.
\end{align}

\section{Parameters for the multi asset model\label{app:calib_het}}
Table \ref{tab:calib_multiasset} shows the parameter values for the model with multiple assets.
\begin{table}
\footnotesize
\begin{center}
\begin{tabular}{ccccccccccccccc}
\toprule 
Parameters & $H$ & $y^h$ & $\psi^h$ & \underline{h} & $\beta$ & \makecell{$\gamma^1$\\ $\gamma^2$} & $\underline{b}$ & \makecell{$B$ \\ $\zeta^b$}& \makecell{$S$ \\ $\zeta^s$} & \makecell{$\zeta^h$ \\ $H^o$} & $\rho$ & $\sigma$ & \makecell{$\mu_1$ \\ $\mu_2$} & d\\
\midrule
Values & 20 & see figure \ref{fig:inc_calib} & see figure \ref{fig:inc_calib} & $5\times10^{-5}$ & 0.867 & \makecell{1\\2} & 0 & \makecell{0.56\\0.25} & \makecell{1\\1} & \makecell{1\\4} & 0.458 & 0.043 & \makecell{$\frac{1}{2}$\\$\frac{1}{2}$} & 0.3\\
\bottomrule
\end{tabular}
\caption{Parameter values chosen for the model with multiple assets.\label{tab:calib_multiasset}}
\end{center}
\end{table}
We continue to model $H = 20$ age groups for each of the two types of risk aversion, \emph{i.e.} a total of 40 households.
The additional parameters, compared to the single asset model, are two levels of risk aversion, as well as the net-supply, borrowing limits and adjustment cost parameters for the new assets.
For both types we continue to choose $u(c)=\frac{c^{1-\gamma}}{1 - \gamma}, v(h^{r}) = \frac{(\underline{h}+h^{r})^{1-\gamma}}{1-\gamma}$, so that the relative risk aversion with respect to the consumption units is the same as the relative risk aversion with respect to the consumption of housing services.
There is a total mass of $\mu_1 = \mu_2 = 0.5$ households for each risk aversion type.
For the risk aversion we choose $\gamma^1 = 1$ for the low risk aversion type and $\gamma^2 = 2$ for the high risk aversion type.
None of the asset can be sold short.
The stock is in unit net supply, $S = 1$, with corresponding adjustment cost parameter $\zeta^s = 1$.
The dividend parameter is chosen to be $d = 0.3$ such that the aggregate dividends amount to 30\% of the aggregate labor income.
The housing supply is given by $H^o = 1$, and $H^{ex} = 0$, \emph{i.e.} all houses are owned by households.
The adjustment cost parameter for house ownership is chosen to be $\zeta^{h} = 4$.
Bonds have a net supply of $0.56$ with associated adjustment cost parameter $\zeta^b = 0.25$.
The age-dependent preference parameter for housing services, $\psi^h$, as well as the parameters governing the evolution of the exogenous shock remain the same as in the single asset model.
\end{appendices}
\end{document}